\documentclass[10pt,a4paper]{article}
\usepackage[utf8]{inputenc}
\usepackage[T1]{fontenc}
\usepackage{amsmath}
\usepackage{amsfonts}
\usepackage{amssymb}
\usepackage{graphicx}
\usepackage{subfig}
\usepackage{cite}
\usepackage[cm]{fullpage}
\usepackage{layout}

\providecommand{\keywords}[1]{{\textit{Keywords}} #1}

\newcommand{\uhat}{\hat{u}}
\newcommand{\vhat}{\hat{v}}
\newcommand{\tuext}{\tilde{u}^\text{ext}}
\newcommand{\tvext}{\tilde{v}^\text{ext}}
\newcommand{\uext}{u^\text{ext}}
\newcommand{\vext}{v^\text{ext}}
\newcommand{\iph}{i+\frac{1}{2}}
\newcommand{\jph}{j+\frac{1}{2}}
\newcommand{\imh}{i-\frac{1}{2}}
\newcommand{\jmh}{j-\frac{1}{2}}
\newcommand{\ipoh}{i+\frac{3}{2}}

\newcommand{\uvec}{\mathbf{U}}
\newcommand{\Uvec}[1]{ \left(
		u^{#1},		
		v^{#1},		
		p^{#1}
\right)}
\newcommand{\Uvect}[1]{\begin{pmatrix}
	u^{#1}\\		
	v^{#1}\\	
	p^{#1}
\end{pmatrix}}
                                                                                                               
\newcommand{\normal}{\mathbf{n}}
\newcommand{\tanv}{ {\boldsymbol{\tau} }}

\newcommand{\lap}{\bigtriangleup}
\newcommand{\xy}[2]{(x_{#1},y_{#2})}
\newcommand\norm[1]{\left\lVert#1\right\rVert}
\begin{document}

\title{Fully implicit and accurate treatment of jump conditions for two-phase incompressible Navier--Stokes equation}
\author{Hyuntae Cho and Myungjoo Kang}
\maketitle
\begin{abstract}
	We present a numerical method for two-phase incompressible Navier--Stokes equation with jump discontinuity in the normal component of the stress tensor and in the material properties. Although the proposed method is only first-order accurate, it does capture discontinuity sharply, not neglecting nor omitting any component of the jump condition. Discontinuities in velocity gradient and pressure are expressed using a linear combination of singular force and tangential derivatives of velocities to handle jump conditions in a fully implicit manner. The linear system for the divergence of the stress tensor is constructed in the framework of the ghost fluid method, and the resulting saddle-point system is solved via an iterative procedure. Numerical results support the inference that the proposed method converges in $L^\infty$ norms even when velocities and pressures are not smooth across the interface and can handle a large density ratio that is likely to appear in a real-world simulation.
	\\
\\
\keywords{Two-phase flows, 	Incompressible Navier-Stokes, Finite difference method, Level-set method}
\end{abstract}

\section{Introduction}
Two-phase incompressible flows are ubiquitous in real life, and thus numerical simulations of these flows are crucial in applications such as oil--water core annular flow, biofluid dynamics, and analysis of gas bubbles rising in water. One popular approach to handling the interface between the two fluids involved is to use a body-fitted mesh. However, in this study, we will narrow our interest to the Cartesian grid method, which is free of mesh generation. The very early work of Peskin \cite{peskin1972flow}, which is also known as the immersed boundary method, is a typical example of a Cartesian grid method. It uses numerical $\delta$-functions to handle singular forces on the interface between fluid and solid. This idea was utilized together with the front-tracking method \cite{unverdi1992front,tryggvason2001front} and the level-set method \cite{sussman1994level,sussman1998improved} for the numerical simulation of incompressible two-phase flows. However, such an approach using smoothed $\delta$-functions cannot avoid numerical diffusion near the interfaces and eventually encounters interfaces with non-zero thicknesses. 

To avoid the smearing out of the interfaces involved in these incompressible two-phase flows, sharp capturing methods have been developed by researchers over the years. For example, the immersed interface method (IIM) \cite{leveque1994immersed} has earned a reputation as a second-order finite difference method for elliptic interface problems. The fundamental idea of IIM has been applied to the incompressible Stokes equation \cite{leveque1997immersed} and Navier--Stokes equation \cite{li2001immersed,lee2003immersed}, on the assumption that the viscosities and densities of the two fluids are identical. Later, introducing augmented variables together with the interfaces, Li et al. \cite{li2007augmented} used IIM to solve Stokes equations with discontinuous viscosities. However, instead of the marker-and-cell (MAC) grid, a collocated grid was used, which caused periodic boundary conditions to be imposed for the pressure. To address this issue, solution methods utilizing the MAC grid have been developed for the two-phase Stokes equation \cite{tan2011implementation,chen2018direct}. For Navier--Stokes equations with discontinuous viscosities, IIM has also demonstrated its success in capturing non-smooth velocities and pressures \cite{tan2008immersed,tan2009immersed}, but relatively less work is done when the density is discontinuous across the interface. (For more detailed explanations and applications of IIM, see \cite{li2006immersed}.)

Another famous sharp capturing method is the ghost fluid method (GFM), which was first introduced by Fedkiw et al. \cite{fedkiw1999non} for capturing contact discontinuity in compressible flows. Its concept was later employed in solving elliptic interface problems \cite{liu2000boundary}. For incompressible flows \cite{kang2000boundary}, techniques by \cite{liu2000boundary} have been applied in approximating viscous terms and solving Poisson's equations of pressure from the projection method \cite{chorin1968numerical}, resulting in the sharp capture of the surface tension. Whereas the viscous term was discretized explicitly in GFM, Sussman et al. \cite{sussman2007sharp} introduced sharp capturing methods that involve semi-implicit treatments of the viscous term, allowing for larger time steps. Even though the details of the implementations by \cite{kang2000boundary} and \cite{sussman2007sharp} differ, Lalanne et al. \cite{lalanne2015computation} have demonstrated that the two approaches are actually equivalent.

Two pioneering works on simulating incompressible flows in the framework of the level-set/ghost fluid method \cite{kang2000boundary,sussman2007sharp} used the projection method to solve for fluid velocity. However, it should be noted that when the projection method is used, the jump conditions of intermediate or predictor velocities differ from those of the original velocities. Nevertheless, the same jump conditions are applied occasionally to simulate two-phase flows. To impose jump conditions accurately, approaches that are alternative to the projection method have been created. One example is the virtual node method, which was first developed for elliptic interface problems \cite{bedrossian2010second,hellrung2012second} and then extended to incompressible flows \cite{assencio2013second,schroeder2014second}. It directly discretizes the Navier--Stokes equation together with the divergence-free condition to obtain a saddle-point linear system of velocity and pressure. On the other hand, Saye has used the gauge method to simulate incompressible flows \cite{saye2017implicit} where the jump conditions were reformulated with auxiliary and gauge variables. Recently, in \cite{theillard2019sharp}, a modified projection method was developed for a sharp capturing method that uses the quad/octree grid. The projection step was repeated until the corrected velocity and pressure satisfied the jump conditions. The researchers also remarked that many existing numeric methods omit some parts of the jump conditions to simplify implementation. 

In this paper, we introduce a new sharp capturing method for two-phase flows, characterized by the accurate and fully implicit treatment of jump conditions. Different from \cite{theillard2019sharp,saye2017implicit}, our method shares similarities with the virtual node method. That is, instead of introducing an auxiliary variable, our method discretizes the divergence of the stress tensor directly in the framework of GFM. Expressions for the jump conditions of velocity gradient and pressure using tangential derivatives of velocities are calculated to develop a ghost fluid method for viscous terms and gradients of pressure. We note that the proposed method is only first-order accurate but considers jump conditions accurately and implicitly. We begin in section \ref{sec:gov_eq} with equations on two-phase incompressible flows and on the movements of the interfaces. Section \ref{section:Numerical Methods} explains the details of the proposed method. Section \ref{sec:Numerical experiment} then follows with descriptions and accounts of the numerical experiments.

\section{Governing Equations}\label{sec:gov_eq}
We consider two incompressible, immiscible, and viscous flows on \(\Omega\), 
\begin{equation}\label{eq:NS}
\begin{gathered}
\rho^\pm (\uvec + \uvec \cdot \nabla \uvec)= \mu^\pm \lap \uvec -\nabla p +\mathbf{f} \text{ on } \Omega^\pm,\\
\nabla \cdot \uvec =0 \text{ on } \Omega.
\end{gathered}
\end{equation}
In addition to \eqref{eq:NS}, jump conditions are given at the interface \(\Gamma\):
\begin{equation}\label{eq:jmp_condition}
\begin{gathered}
\left[\uvec\right]=0,\\
\left[ \sigma \normal\right]=\mathbf{G},
\end{gathered}
\end{equation}
for stress tensor \(\sigma=\mu \left(\nabla\uvec +\nabla\uvec ^T\right) - p\mathbf{I} \). $\left[\mathbf{V}\right]= \mathbf{V}^+- \mathbf{V}^-$ denotes the jump condition along the interface, where the superscripts "$+$" and "$-$" refer to $\Omega^\pm$. Here, \(\normal \) is a normal to the interface, and $\mathbf{G}$ is a singular force term across the interface. We are especially interested in the case where $\mathbf{f}= -\rho \mathbf{g}$ and $\mathbf{G}=\beta \kappa \normal$, where $\mathbf{g}$ refers to gravity, $\kappa$ denotes the mean curvature of the interface, and $\beta$ is a surface tension coefficient. 

In this study, the level-set method \cite{osher1988fronts} is used to capture the interface of two different fluids as a zero level-set of the continuous function $\phi$. Thus, the interface and two sub-domains at time $t$ can be represented as
\begin{equation}\label{eq:interface_position}
\begin{gathered}
\Gamma =\{\mathbf{x} \in \Omega | \phi(\mathbf{x}, t)=0\},\\
\Omega^+ =\{\mathbf{x} \in \Omega | \phi(\mathbf{x}, t)>0\},\\
\Omega^- =\{\mathbf{x} \in \Omega | \phi(\mathbf{x}, t)<0\}.
\end{gathered}	
\end{equation}
An advantage of the level-set method is that the evolution of the interface $\Gamma$ with the fluid velocity $\uvec$ can be formulated as 
\begin{equation}\label{eq:LV_advec}
\phi_t +\uvec \cdot \nabla \phi =0.
\end{equation}
Furthermore, geometric quantities of the interfaces, such as the normal $\normal$ and curvature $\kappa$ in \eqref{eq:jmp_condition}, are calculated using the level-set function:
\begin{equation}\label{eq:geometric_quantities}
\begin{aligned}
\normal &= \frac{\nabla \phi}{|\nabla \phi|},\\
\kappa &= \nabla \cdot \normal.
\end{aligned}
\end{equation}
For a more detailed explanation and application of the level-set method, see \cite{osher2004level,gibou2018review}.

\section{Numerical Methods} \label{section:Numerical Methods}

\begin{figure}
	\centering
	\includegraphics[width=0.5\textwidth]{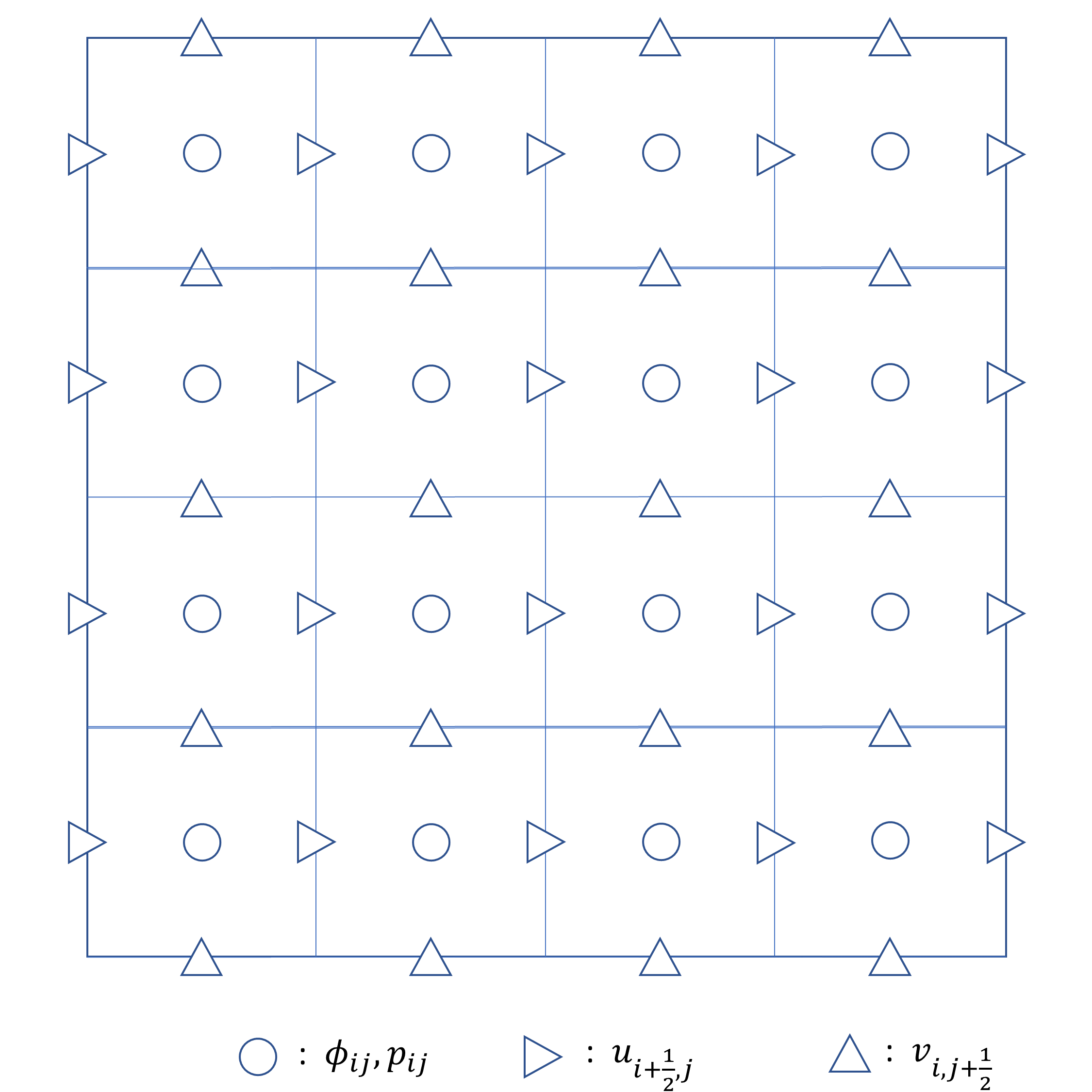}
	\caption{MAC grid}
	\label{fig:MAC}
\end{figure}
We use a staggered MAC grid\cite{harlow1965numerical} for spatial discretization. Pressure $p$ and level-set function $\phi$ are placed in the cell centers, whereas the values from $\uvec$ are stored on the cell faces. Figure \ref{fig:MAC} illustrates the locations of the variables in 2D.

Overall discretization of the proposed method is similar to that of \cite{schroeder2014second}. At the beginning of each time step, $\phi^n$ is advected to $\phi^{n+1}$ via the discretization  of \eqref{eq:LV_advec} with third-order total variation diminishing (TVD) Runge-Kutta \cite{shu1988efficient} in time and fifth-order weighted essentially non-oscillatory (WENO) scheme \cite{jiang1996efficient} in space. To discretize \eqref{eq:LV_advec}, the velocity at cell center needs to be defined, which will be done later. After the advection, $\phi^{n+1}$ is reinitialized through the solving of the partial differential equation
\begin{equation}\label{eq:reinitialization}
\phi_\xi +\text{sign}(\phi)(|\nabla \phi|-1)=0
\end{equation} 
for pseudo time $\xi$. Discretization of \eqref{eq:reinitialization} follows that of \cite{min2007second}, which uses second-order essentially non-oscillatory (ENO) scheme and TVD-RK2 with subcell resolution technique. The $\phi^{n+1}$ on the cell faces are evaluated via linear interpolation of the value defined at the cell center:
\begin{equation*}
\begin{aligned}
\phi^{n+1}_{i+\frac{1}{2},j}&= \frac{\phi^{n+1}_{i,j}+\phi^{n+1}_{i+1,j}}{2}, \\
\phi^{n+1}_{i,j+\frac{1}{2}}&= \frac{\phi^{n+1}_{i,j}+\phi^{n+1}_{i,j+1}}{2}. \\
\end{aligned}
\end{equation*}
These values will determine if the center of a cell face belongs to either $\Omega^+$ or $\Omega^-$. For the Navier--Stokes equation, we use semi-Lagrangian with backward difference formula to construct a saddle-point system of $\uvec$ and $p$:
\begin{equation}\label{eq:SLBDF}
\begin{gathered}
\rho^{n+1} \frac{\uvec^{n+1}-\uvec_d^n}{\Delta t} = \left(\mu \lap \uvec -\nabla p +\mathbf{f}\right)^{n+1},\\
\left(\nabla \cdot \uvec\right)^{n+1}=0.
\end{gathered}
\end{equation}
Here, $\uvec_d^n$ is an approximation of $\uvec^n$ at departure point $\mathbf{x}_d$ traced backward along the characteristic curve $\frac{d \mathbf{x}}{d t}=\uvec$ from time level $t^{n+1}$. Geometric quantities \eqref{eq:geometric_quantities} that appear in the jump condition are approximated with $\phi^{n+1}$ using second-order central differences. Because the Navier--Stokes equation is implicitly discretized, whether the grid point belongs to $\Omega^+$ or $\Omega^-$ is determined by the sign of $\phi^{n+1}$ rather than that of $\phi^n$.

Discretizing viscous terms and pressure using jump condition \eqref{eq:jmp_condition} is quite a challenging problem. When the viscous term is treated explicitly, the higher truncation error near the interface does not spread out. Furthermore, as the grid is refined, a significant restriction is imposed on the time step. On the other hand, if the viscous term is discretized implicitly using the projection method, the jump condition of $\uvec$ will differ from the jump condition of the intermediate or predictor velocity. However, this difference is sometimes ignored. Direct discretization of Navier--Stokes equation \eqref{eq:SLBDF} enables us to treat the viscous term implicitly with accurate jump conditions.

In the following subsections, we will first derive a jump condition equivalent to \eqref{eq:jmp_condition} without neglecting any component. To emphasize the sharp discretization of viscous terms and pressure using the jump condition, a solution method for two-phase steady-state Stokes equations that involve these jump conditions will be presented. Afterward, with some modifications, the numerical algorithm for \eqref{eq:SLBDF} will be completed.

\subsection{Derivation of jump conditions}
A jump condition equivalent to \eqref{eq:jmp_condition} is first derived through the replacement of normal derivatives with a linear combination of Cartesian and tangential derivatives, as was done for the elliptic interface problem in \cite{cho2019second,egan2020XGFM}. The normal and tangent vectors of the interface are then defined as \(\normal =(n_x,n_y)\) and $\tanv=(-n_y, n_x)$, respectively. 

First, the inner product of \eqref{eq:jmp_condition} and \(\tanv\) is calculated:
\begin{equation} \label{eq:jmp_cond_innerp_tangent}
\begin{aligned}
\mathbf{G} \cdot \tanv &=[\mu (\tanv \cdot \nabla \uvec  \cdot \normal + \normal\cdot \nabla \uvec \cdot \tanv)]\\
&= [\mu (-n_y u_\normal+n_x v_\normal + n_x  u_\tanv + n_y v_\tanv)].
\end{aligned}
\end{equation}
Here, \(u_\normal =\nabla u \cdot \normal , u_\tanv =\nabla u\cdot \tanv , v_\normal =\nabla v \cdot \normal , v_\tanv =\nabla v\cdot \tanv\).
Because divergence is invariant under rotation,
\begin{equation}\label{eq:div_free_coordinate}
n_x  u_\normal + n_y v_\normal -n_y u_\tanv + n_x v_\tanv= \frac{\partial \left(\uvec \cdot \normal\right)}{\partial\normal} +\frac{\partial\left( \uvec \cdot \tanv\right)}{\partial \tanv}=\nabla \cdot \uvec =0.
\end{equation}
\(n_y\) is multiplied to \eqref{eq:jmp_cond_innerp_tangent}, and \(n_y v_\normal\) is substituted for \(-n_x u_\normal +n_y u_\tanv -n_x v_\tanv \):
\begin{equation*}
n_y \mathbf{G}\cdot \tanv = \left[ \mu (-u_\normal + 2 n_x n_y u_\tanv +(n_y^2 - n_x ^2 ) v_\tanv)   \right].
\end{equation*}
Similarly, the following equation is obtained:
\begin{equation*}
n_x \mathbf{G}\cdot \tanv = \left[ \mu (v_\normal + 2 n_x n_y v_\tanv +(n_x^2 - n_y ^2 ) u_\tanv)   \right].
\end{equation*}
The jump conditions of the derivatives in the Cartesian directions can be decomposed into conditions in the normal and tangential directions:
\begin{align*}
[\mu u_x ] &= [\mu u_\normal] n_x - [\mu u_\tanv ] n_y \\
&= [\mu u_\normal] n_x - [\mu ]u_\tanv n_y \\
[\mu u_y ] &= [\mu u_\normal] n_y + [\mu u_\tanv ] n_x \\
&= [\mu u_\normal] n_y + [\mu ]u_\tanv n_x .
\end{align*}
The jump condition \([u]=[v]=0\) is used, resulting in \([\mu u_\tanv]=[\mu]u_\tanv^- +\mu^+ [u_\tanv] = [\mu] u_\tanv \). The superscripts "$+$" and "$-$" for $u_\tau$ are dropped because $\left[u_\tau\right]=0$. When these resulting expressions are combined, the jump condition for the Cartesian component of $\left[\mu\nabla u\right]$ is produced:
\begin{equation}\label{eq:formul_jmp_cond_u}
\begin{aligned}
\left[\mu u_x \right]&=  [\mu](2n_x^2 n_y -n_y) u_\tanv  + [\mu](n_x n_y^2 -n_x^3) v_\tanv - n_x n_y \mathbf{G} \cdot \tanv,\\
[\mu u_y]&= [\mu](2n_x n_y^2 + n_x) u_\tanv  + [\mu]( n_y^3 -n_x^2n_y) v_\tanv - n_y^2 \mathbf{G} \cdot \tanv.
\end{aligned}
\end{equation}
Similarly, the formula for \([\mu \nabla v]\) can be derived:
\begin{equation}\label{eq:formul_jmp_cond_v}
\begin{aligned}
\left[\mu v_x\right] &=  [\mu](-2n_x^2 n_y -n_y) v_\tanv  + [\mu](n_x n_y^2 -n_x^3) u_\tanv + n_x^2 \mathbf{G} \cdot \tanv,\\
[\mu v_y] &= [\mu](-2n_x n_y^2 + n_x) v_\tanv  + [\mu]( n_y^3 -n_x^2 n_y) u_\tanv + n_x n_y \mathbf{G} \cdot \tanv.
\end{aligned}
\end{equation}
Lastly, the inner product of \eqref{eq:jmp_condition} and \(\normal\) is calculated:
\begin{align*}
\mathbf{G} \cdot \normal &=[2\mu (\normal \cdot \nabla \uvec  \cdot \normal ) -p].
\end{align*}
From \eqref{eq:div_free_coordinate},
\begin{align*}
\normal \cdot \nabla \uvec \cdot \normal =-\tanv \cdot \nabla \uvec \cdot \tanv=-n_y u_\tanv +n_x v_\tanv.
\end{align*}
Therefore,
\begin{equation}\label{eq:formul_jmp_cond_p}
[p]= -2[\mu](-n_y u_\tanv +n_x v_\tanv)-\mathbf{G} \cdot \normal .
\end{equation}
Through this formula, we express the jump condition $\left[\mu \nabla \uvec\right],\left[p\right]$ as a linear combination of the singular force $\mathbf{G}$ and the tangential derivatives of the velocities. 

\paragraph{Remark} The jump condition formula can be extended to three dimensions with few modifications. After constructing two tangent vectors with respect to the normal vector, one can derive two jump conditions of velocities, similar to \eqref{eq:jmp_cond_innerp_tangent}, and obtain a divergence-free condition in the normal and tangent coordinates. With an appropriate linear combination of these three equations, jump conditions for the velocity gradients can be obtained. For the jump condition of pressure, we first calculate the inner product of the normal vector and \eqref{eq:jmp_condition}. After replacing the normal derivatives of the velocities with the tangential derivatives of the velocities using divergence-free condition, we obtain a three-dimensional version of \eqref{eq:formul_jmp_cond_p}.

\subsection{Numerical methods for two-phase steady-state Stokes equation}\label{section:NumericMethods_XGFM_STOKES}

Ignoring the material derivative with density in \eqref{eq:NS}, we consider the incompressible two-phase steady-state Stokes equation:
\begin{equation}\label{eq:Stokes1}
\begin{gathered}
\mu^{\pm} \lap \uvec-\nabla p + \mathbf{f}=0 \text{ on } \Omega^{\pm},\\
\nabla \cdot \uvec=0 \text{ on } \Omega ,\\
\left[\sigma \normal \right]= G \text{ on } \Gamma.
\end{gathered}
\end{equation}
Here, we introduce numerical methods to solve \eqref{eq:Stokes1} in a sharp manner. The idea of xGFM\cite{egan2020XGFM} is extended to jump conditions where $\uvec$ and $p$ are coupled together. We construct an iterative method that corrects the jump conditions and solutions together for every iterative step. Details of the algorithms will be presented under the assumption of two dimensions, whereas extension to three dimensions will be straightforward.
\subsubsection{Visiting two-phase steady-state Stokes equation with ghost fluid method}\label{subsec:stokes_gfm_default}
Under the assumption of two dimensions, the following are considered:
\begin{equation}\label{eq:Stokes_equation}
\begin{gathered}
\mu^\pm \lap u -p_x = -f_1 \text{ on } \Omega^\pm \\
\mu^\pm \lap v -p_y = -f_2 \text{ on } \Omega^\pm\\
u_x+v_y= 0 \text{ on } \Omega^\pm
\end{gathered}
\end{equation}
with the jump conditions 
\begin{equation}\label{eq:jmp_cartesian}
\begin{aligned}
\left[\mu \nabla u\right]&= \mathbf{c} \text{ on } \Gamma\\
[\mu \nabla v]&= \mathbf{d} \text{ on } \Gamma\\
[p]&= a \text{ on } \Gamma
\end{aligned}
\end{equation}
instead of \eqref{eq:jmp_condition}. Although \eqref{eq:Stokes_equation} with jump conditions \eqref{eq:jmp_cartesian} result in over-determined partial differential equations, GFM is a suitable method for solving such a problem. When the methodologies of GFM are followed, \eqref{eq:Stokes_equation} with \eqref{eq:jmp_cartesian} may be discretized as follows:
\begin{equation}\label{eq:discretization_stoeks}
\begin{aligned}
(\mu \lap u)_{\iph,j}^\text{GFM} - \frac{p_{i+1,j}-p_{i,j}}{\Delta x} &=-f_1 \xy{\iph}{j} +c_{u} +a_u \\
(\mu \lap v)_{i,\jph}^\text{GFM} - \frac{p_{i,j+1}-p_{i,j}}{\Delta y} &=-f_2 \xy{i}{\jph} +d_{v} +a_v \\
\frac{  u_{\iph,j}-u_{\imh,j}}{\Delta x} + \frac{v_{i,\jph}-v_{i,\jmh}}{\Delta y}&=0.
\end{aligned}
\end{equation}
Here, $\mu\lap u$ and $p_x$ at $\xy{\iph}{j}$ are approximated as $(\mu \lap u)_{\iph,j}^\text{GFM}-c_u$ and $ \frac{p_{i+1,j}-p_{i,j}}{h} + a_u$, respectively. $(\mu \lap u)_{\iph,j}^\text{GFM}$ denotes the discrete Laplacian that appears in GFM\cite{liu2000boundary}, and $c_u$ is the correction term added to the right-hand side. If $\xy{\iph}{j} \in \Omega^\pm$ is assumed, 
\begin{equation*}
\begin{aligned}
(\mu \lap u)_{\iph,j}^\text{GFM} =& \left.\left( \hat{\mu}_{R}( u_{\ipoh,j}-u_{\iph,j})+\hat{\mu}_{L}( u_{\imh,j}-u_{\iph,j})\right)\middle/ \Delta x^2 \right.+ \\
&\left.\left(\hat{\mu}_{T}( u_{\iph,j+1}-u_{\iph,j}) +\hat{\mu}_{B}( u_{\iph,j-1}-u_{\iph,j})\right)\middle/ \Delta  y^2\right.,\\
c_u = & c_R+c_L +c_T+ c_B,
\end{aligned}
\end{equation*}
where \begin{equation*}
\hat{\mu}_R = 
\begin{cases} 
\mu^\pm &\text{ if } \phi_{\iph,j}\phi_{\ipoh,j}>0)\\
\frac{\mu^+ \mu^-}{\mu^\pm \theta_R+ \mu^\mp (1-\theta_R)} &\text{ if } \phi_{\iph,j}\phi_{\ipoh,j}\leq 0)
\end{cases}
\end{equation*}
and 
\begin{equation*}
c_R = 
\begin{cases} 
0 &\text{ if } \phi_{\iph,j}\phi_{\ipoh,j}>0)\\
\pm \hat{\mu}_R \frac{(1-\theta_R)[\mu u_x]_R
}{\mu^\mp \Delta x} &\text{ if } \phi_{\iph,j}\phi_{\ipoh,j}\leq 0)
\end{cases}
\end{equation*}
for \(\theta_R=\frac{|\phi_{\iph,j}|}{|\phi_{\iph,j}|+|\phi_{i+\frac{3}{2},j}|}\).
$\hat{\mu}_L,\hat{\mu}_T,\hat{\mu}_B$ and $c_L,c_T,c_B$ are defined similarly. (For a more detailed explanation, see \cite{liu2000boundary}.) In a similar fashion, the correction term for $p_x$ is determined as
\begin{equation*}
a_u = a_L+a_R,
\end{equation*}
where
\begin{align*}
a_R &= \begin{cases}
0 &\text{ if } \phi_{\iph,j} \phi_{i+1,j}>0\\
\mp\frac{ [p]_R}{\Delta x} &\text{ if } \phi_{\iph,j} \phi_{i+1,j}<0
\end{cases},\\
a_L &= \begin{cases}
0 &\text{ if } \phi_{\iph,j} \phi_{i,j}>0\\
\pm\frac{ [p]_L}{\Delta x} &\text{ if } \phi_{\iph,j} \phi_{i,j}<0
\end{cases}.
\end{align*}
The formulation of $(\mu \lap v)_{i,\jph}^\text{GFM},d_v,a_v$ is similar to that of $(\mu \lap u)_{\iph,j}^\text{GFM},c_u,a_u$. For simplicity, \eqref{eq:discretization_stoeks} is rewritten as 
\begin{equation}\label{eq:discretization_stokes_simple}
A(u,v,p)=b(a,\mathbf{c},\mathbf{d},\mathbf{f}).
\end{equation}
Notably, $A$ is multi-linear with respect to $u, v,$ and $p$, whereas \(b\) is multi-linear with respect to \(a, \mathbf{c},\) and \(\mathbf{d}\).
\paragraph{Remark} Note that linear system \eqref{eq:discretization_stoeks} is symmetric. A correction term is not added to the divergence-free equation \(u_x + v_y=0\). Furthermore, a $O(1)$ truncation error occurs near the interface. If one uses jump conditions $[\mu u_x ],[ \mu v_y]$ to discretize a divergence-free equation, the truncation error near the interface becomes $O(h)$ for $h=\max(\Delta x, \Delta y)$ but breaks the symmetry of the matrix when $\mu^-$ and $\mu^+$ are different. Because $\mu \lap \uvec -\nabla p +\mathbf{f}=0$ are discretized with $O(1)$ truncation error near the interface, considering the jump condition for a divergence-free equation will not increase the order of convergence dramatically, but rather, will make the linear system difficult to solve by creating a non-symmetric linear system.

\subsubsection{Velocity extrapolation algorithm and tangential derivative at the interface}\label{subsec:velocity_extrapolation}
Here, the velocity extrapolation algorithm of \cite{egan2020XGFM} is revisited. The following pseudo time-dependent partial differential equation is first considered:
\begin{equation}\label{eq:psuedo_extra}
\begin{gathered}
\frac{\partial \hat{u}}{\partial t} +\text{sign}(\phi)\normal \cdot \nabla \hat{u}=0,\\
\frac{\partial \hat{v}}{\partial t} +\text{sign}(\phi)\normal \cdot \nabla \hat{v}=0,\\
\hat{u}=u, \hat{v} =v \text{ on } \Gamma.
\end{gathered}
\end{equation}
The steady-state solution of \eqref{eq:psuedo_extra} can be viewed as an extrapolation of $u,v$ off the interface and will be used to approximate tangential derivatives of $u,v$ on the interface. Equation \eqref{eq:psuedo_extra} is then discretized using a first-order upwind scheme. For example, if $\phi_{\iph,j}>0$ is assumed, it is discretized as
\begin{align} \label{eq:one_timestep_extrapolation}
\frac{\uhat^{l+1}_{\iph,j} - \uhat^l_{\iph,j}}{\Delta t_{\iph,j}} + (n_x^+ D_x^- \uhat + n_x^- D_x^+ \uhat )+(n_y^+ D_y^- \uhat + n_y^- D_y^+ \uhat )=0
\end{align}
for $n_x^+ = \max(n_x, 0), n_x^- = \min(n_x,0)$, and similarly defined $n_y^\pm$. The boundary condition of \eqref{eq:psuedo_extra} is applied on $\Gamma$ using a sub-cell resolution technique:
\begin{equation} \label{eq:extrapolation_boundary}
\begin{aligned}
D_x^- \uhat_{\iph,j} = \begin{cases}
\frac{\uhat^l_{\iph,j}-\uhat^l_{\imh,j}}{\Delta x} &\text{ if } \phi_{\iph,j}\phi_{\imh,j}>0 \\
\frac{\uhat^l_{\iph,j}-u_\Gamma}{\theta_L \Delta x} &\text{ if } \phi_{\iph,j}\phi_{\imh,j}<0 
\end{cases}.
\end{aligned}
\end{equation}
\(\theta_L = \frac{|\phi_{\iph,j}|}{|\phi_{\iph,j}|+|\phi_{\imh,j}|}\) is measured to approximate the location of the interface. The interfacial value $u_\Gamma$ is obtained according to the formula in \cite{liu2000boundary}:\begin{equation*}
u_\Gamma = \left.\left(\mu_2 \theta_L u_{\iph,j}+\mu_1 (1-\theta_L) u_{i-\frac{1}{2},j} -\text{sign} (\phi_{\iph,j})  \left(\left[\mu u_x \right](1-\theta_L)\theta_L \Delta x \right)\right)\middle/ \hat{\mu}\right.
\end{equation*} for 
\begin{equation*}
\mu_1 = \begin{cases}
\mu^+ \text{ if } \phi_{\iph,j}>0\\\mu^- \text{ if } \phi_{\iph,j}<0
\end{cases},\quad\mu_2 = \begin{cases}
\mu^+ \text{ if } \phi_{i-\frac{1}{2},j}>0\\\mu^- \text{ if } \phi_{i-\frac{1}{2},j}<0
\end{cases}, \quad \hat{\mu}=\mu_2 \theta_L+ \mu_1 (1-\theta_L).
\end{equation*}
Other derivatives and interface boundary conditions are computed similarly. To avoid small time-stepping in the whole domain, the local pseudo time step is set to $\Delta t_{\iph,j}= CFL\times \min(\theta_R\Delta x,\theta_L\Delta x, \theta_B\Delta y,\theta_T\Delta y)$, where $\theta_L=1$ if $\phi_{\iph,j}\phi_{\imh,j}>0$. The idea of taking a grid-dependent time step and using it in reinitialization can be found in \cite{min2010reinit}.

Equation \eqref{eq:psuedo_extra} is solved only on the grid points near the interface, and $u^\text{ext},v^\text{ext}$ is denoted as the steady-state solution of \eqref{eq:psuedo_extra}. Specifically, extrapolations were performed for the points where $|\phi_{\iph,j}|,|\phi_{i,\jph}|\leq 5\max(\Delta x , \Delta y)$ and $u^\text{ext}_{\iph,j} =\uhat^{l+1}_{\iph,j}$ when $ |\uhat^{l+1}_{\iph,j}- \uhat^l_{\iph,j}|<\epsilon$ for every $i,j$ with tolerance $\epsilon$. After steady-state solutions $u^\text{ext}$ and $v^\text{ext}$ are obtained, the followings are defined: 
\begin{equation}
\uext_\tanv = \nabla \uext \cdot \tanv, \quad \vext_\tanv = \nabla \vext \cdot \tanv
\end{equation} 
at grid points $\xy{\iph}{j}$ for
\begin{align*}
\nabla u^\text{ext} \xy{\iph}{j} = \begin{pmatrix}
\frac{u^\text{ext}_{\ipoh,j} -\uext_{\imh,j}}{2\Delta x}\\
\frac{u^\text{ext}_{i,j+1} -\uext_{i,j-1}}{2\Delta y}
\end{pmatrix}, \quad  \nabla v^\text{ext} \xy{\iph}{j} = \begin{pmatrix}
\frac{v^\text{ext}_{i+1,\jph}+\vext_{i+1,\jmh} -\vext_{i,\jph}-\vext_{i,\jmh}}{2\Delta x}\\
\frac{v^\text{ext}_{i+1,\jph}-\vext_{i+1,\jmh} +\vext_{i,\jph}-\vext_{i,\jmh}}{2\Delta y}
\end{pmatrix}.
\end{align*}
$\tanv_{\iph,j}=(-n_y, n_x)_{\iph,j}$ is obtained from $\normal=(n_x,n_y)_{\iph,j}$ computed with the level-set function $\phi$. Because the steady-state solution $u^\text{ext},v^\text{ext}$ is a reasonable extrapolation of \(u,v\) off the interface, $u^\text{ext}_\tanv,v^\text{ext}_\tanv$ can be viewed as an extension of $u_\tanv, v_\tanv$ at $\Gamma$ to the grid points near the interface. $\uext_\tanv,\vext_\tanv$ at $\xy{i}{\jph}$ can be computed with few modifications. Note that $\text{sign}(\phi) \normal$ does not depend on $\uhat,\vhat$, whereas $\uhat^{l+1}_{\iph,j}$ linearly depends on $\uhat^{l}_{\iph,j}$ and $u_\Gamma$. Therefore, we may conclude that $u^\text{ext}_\tanv$ is a linear combination of $\left[\mu u_x\right],\left[\mu u_y\right]$, and $u$.

\subsubsection{Iterative procedure}\label{subsec:Stokes_iterative}
Steady-state Stokes equations \eqref{eq:Stokes_equation} with jump condition \eqref{eq:jmp_condition} can be reformulated into equivalent systems of partial differential equations using \eqref{eq:formul_jmp_cond_u}, \eqref{eq:formul_jmp_cond_v}, and \eqref{eq:formul_jmp_cond_p} to develop iterative methods:
\begin{equation}\label{eq:reformulation_stokes}
\begin{gathered}
\mu \lap u -p_x = -f_1 \text{ on } \Omega^\pm\\
\mu \lap v -p_y = -f_2 \text{ on }  \Omega^\pm\\
u_x+v_y= 0 \text{ on } \Omega^\pm
\end{gathered}
\end{equation}
with the jump condition
\begin{equation}\label{eq:reformulation_jmp}
\begin{gathered}
\left[\mu \nabla u\right]=\begin{pmatrix}
[\mu](2n_x^2 n_y -n_y) u_\tanv & + [\mu](n_x n_y^2 -n_x^3) v_\tanv& - n_x n_y G \cdot \tanv\\
[\mu](2n_x n_y^2 + n_x) u_\tanv & + [\mu]( n_y^3 -n_x^2n_y) v_\tanv& - n_y^2 G \cdot \tanv
\end{pmatrix} \text{ on } \Gamma\\
[\mu\nabla v] =\begin{pmatrix}
[\mu](-2n_x^2 n_y -n_y) v_\tanv&  + [\mu](n_x n_y^2 -n_x^3) u_\tanv& + n_x^2 G \cdot \tanv\\
[\mu](-2n_x n_y^2 + n_x) v_\tanv&  + [\mu]( n_y^3 -n_x^2 n_y) u_\tanv& + n_x n_y G \cdot \tanv
\end{pmatrix} \text{ on } \Gamma\\
[p]= -2[\mu](-n_y u_\tanv +n_x v_\tanv)-G \cdot \normal \text{ on } \Gamma.
\end{gathered}
\end{equation}
Equation \eqref{eq:reformulation_stokes} is discretized according to \ref{subsec:stokes_gfm_default}, where jump condition \eqref{eq:reformulation_jmp} is discretized through the setting of $u_\tanv =\uext_\tanv,v_\tanv= \vext_\tanv$ obtained from \ref{subsec:velocity_extrapolation}. Although the discretized equation is linear, the solution of \eqref{eq:reformulation_stokes} depends on the jump condition, whereas jump condition \eqref{eq:reformulation_jmp} depends on the solution, which makes the linear system difficult to solve. Therefore, the solution of the Stokes equation is determined via the following iterative steps. 

Given \(a^k,\mathbf{c}^k, \mathbf{d}^k\), the value of \((\tilde{u}^{k+1},\tilde{v}^{k+1},\tilde{p}^{k+1})\) is solved through the discretization of the following systems according to \ref{subsec:stokes_gfm_default} :
\begin{equation}\label{eq:stokes_iter}
\begin{aligned}
\mu \lap \tilde{u}^{k+1} -\tilde{p}_x^{k+1} &= -f_1 \text{ on } \Omega^\pm\\
\mu \lap \tilde{v}^{k+1} -\tilde{p}^{k+1}_y &= -f_2 \text{ on }\Omega^\pm\\
\tilde{u}^{k+1}_x+\tilde{v}^{k+1}_y&= 0 \text{ on } \Omega^\pm\\
\left[\mu \nabla \tilde{u}^{k+1}\right]&= \mathbf{c}^{k} \text{ on } \Gamma\\
[\mu \nabla \tilde{v}^{k+1}]&= \mathbf{d}^{k} \text{ on } \Gamma\\
[\tilde{p}^{k+1}]&= a^k \text{ on } \Gamma.
\end{aligned}
\end{equation}
Afterward, via the velocity extrapolation algorithm of \ref{subsec:velocity_extrapolation}, the steady-state solution \( \tilde{u}^\text{ext} ,\tilde{v}^\text{ext}\), with boundary condition given by the interface value of \(\tilde{u}^{k+1}\text{ and }\tilde{v}^{k+1}\), is computed. With the use of the relations \eqref{eq:formul_jmp_cond_u}, \eqref{eq:formul_jmp_cond_v}, and \eqref{eq:formul_jmp_cond_p}, the following equations are defined:
\begin{equation}\label{eq:jump_cond_iterative}
\begin{aligned}
\tilde{\mathbf{c}}^{k+1} =& 
\begin{pmatrix}	
[\mu](2n_x^2 n_y -n_y) \tuext_\tanv& + [\mu](n_x n_y^2 -n_x^3) \tvext_\tanv &- n_x n_y G \cdot \tanv\\
[\mu](2n_x n_y^2 + n_x) \tuext_\tanv& + [\mu]( n_y^3 -n_x^2n_y)\tvext_\tanv& - n_y^2 G \cdot \tanv
\end{pmatrix}\\
\tilde{\mathbf{d}}^{k+1} =& \begin{pmatrix}
[\mu](-2n_x^2 n_y -n_y) \tvext_\tanv&  + [\mu](n_x n_y^2 -n_x^3) \tuext_\tanv&  + n_x^2 G \cdot \tanv\\
[\mu](-2n_x n_y^2 + n_x) \tvext_\tanv&   + [\mu]( n_y^3 -n_x^2 n_y) \tuext_\tanv&  + n_x n_y G \cdot \tanv
\end{pmatrix}\\
\tilde{a}^{k+1}=& -2[\mu](-n_y \tuext_\tanv +n_x \tvext_\tanv)-G \cdot \normal .
\end{aligned}
\end{equation}
$\tilde{\mathbf{c}}^{k+1}$ and $\tilde{\mathbf{d}}^{k+1}$ are defined at grid points $\xy{\iph}{j}$ and $\xy{i}{\jph}$, respectively. 

The residual vectors are then defined as follows:
\begin{equation*}
\begin{aligned}
\tilde{r}^{k+1} &=A(\tilde{u}^{k+1},\tilde{v}^{k+1},\tilde{p}^{k+1})-b(\tilde{a}^{k+1},\tilde{\mathbf{c}}^{k+1},\tilde{\mathbf{d}}^{k+1},\mathbf{f}),\\
{r}^{k}& =A({u}^{k},{v}^{k},{p}^{k})-b({a}^{k},{\mathbf{c}}^{k},{\mathbf{d}}^{k},\mathbf{f}).
\end{aligned}
\end{equation*} 
The value of $\eta_k$ that minimizes the 2-norm of
\begin{equation*}
r^{k+1}= (1-\eta_k)	r^k + \eta_k \tilde{r}^{k+1}
\end{equation*}
is computed, resulting in
\begin{equation}\label{eq:eta_formula}
\eta_k =\frac{{r^k \cdot (r^k -\tilde{r}^{k+1})}}{\norm{\tilde{r}^{k+1}-r^k}_2^2}.
\end{equation}
The $(k+1)$-th iterative solution is defined to be a linear combination of the solutions with weights $1-\eta_k$ and $\eta_k$. For example,
\[u^{k+1}= (1-\eta_k)	u^k + \eta_k \tilde{u}^{k+1}.\]
$v^{k+1},p^{k+1},a^{k+1},\mathbf{c}^{k+1},$ and $\mathbf{d}^{k+1}$ are computed similarly. Because of the multi-linear property of \eqref{eq:discretization_stokes_simple}, the following equation is obtained:
\begin{equation}
{r}^{k+1} =A({u}^{k+1},{v}^{k+1},{p}^{k+1})-b({a}^{k+1},{\mathbf{c}}^{k+1},{\mathbf{d}}^{k+1},\mathbf{f}).
\end{equation}
The iterative procedure is stopped if \(\norm{r^{k+1}}_2\) or \(\norm{ (u^{k+1},v^{k+1},p^{k+1})-(u^k,v^k,p^k) } _\infty \) goes under the threshold. 

\paragraph{Remark}
The minimum residual method (MINRES)\cite{paige1975solution} is used to solve linear system \eqref{eq:stokes_iter} for each iterative step. Because $A$ has both positive and negative eigenvalues, incomplete Cholesky decomposition is not available. For example, given the matrix
\begin{equation} \label{eq:preconditioner}
M=
\begin{pmatrix}
-(\mu \lap)^\text{GFM}  & 0 & 0\\
0 &    -(\mu \lap)^\text{GFM}  & 0\\
0&  0& \alpha I
\end{pmatrix}
\end{equation}
for positive value $\alpha$, $M$ is positive-definite, and thus we set the preconditioner of $A$ for the MINRES method as an incomplete Cholesky decomposition of $M$.

\subsection{Numerical methods for two-phase incompressible Navier--Stokes equation}
We now introduce a new method of discretizing incompressible Navier--Stokes equations using the idea described in \ref{section:NumericMethods_XGFM_STOKES}. As mentioned earlier, the basic framework is a semi-Lagrangian with a backward difference formula. For the Navier--Stokes equation, we assume $\mathbf{f}=-\rho \mathbf{g}$ and $\mathbf{G}= \beta \kappa \normal$.
\begin{equation}\label{eq:discretize_NS_overview}
\begin{aligned}
\rho_{\iph,j}^{n+1}\left(\frac{u_{\iph,j}^{n+1}-u^n_d}{\Delta t}\right)&=\left(\mu\lap u-p_x \right)^{n+1} ,\\
\rho_{i,\jph}^{n+1}\left(\frac{v_{i,\jph}^{n+1}-v^n_d}{\Delta t}\right)&=\left(\mu\lap v-p_y\right)^{n+1} -\rho_{i,\jph}^{n+1}g ,\\
\frac{u^{n+1}_{\iph,j}-u^{n+1}_{\imh,j}}{\Delta x}+&\frac{v^{n+1}_{i,\jph}-v^{n+1}_{i,\jmh}}{\Delta y}=0.\\
\end{aligned}
\end{equation}
One time-step procedure of the proposed Navier--Stokes (NS) solver is:
\begin{enumerate}
	\item Compute the departure point and interpolate $u^{n}_d,v^n_d$.
	\item Choose appropriate $\rho_{\iph,j}^{n+1},\rho_{i,\jph}^{n+1}$, and construct saddle-point system corresponding to \eqref{eq:discretize_NS_overview} under the condition that $[\mu \nabla \uvec]^{n+1}$ is given.
	\item Apply iterative procedure to solve for $\uvec^{n+1}$.
\end{enumerate}

\subsubsection{Discretization of convection term}
The material derivative $\frac{du}{dt}=u_t + \uvec \cdot \nabla u$ is discretized using a semi-Lagrangian method. To approximate the departure point of $\xy{\iph}{j}$, first-order backward integration with linearized velocity is used:
\begin{equation*}
\xy{d}{d}= \xy{\iph}{j}- \Delta t \left(u^n_{\iph,j} ,v^n_{\iph,j}  \right).
\end{equation*}
$v^{n}_{\iph,j}$ is evaluated through a computation of the average $v$ from the nearby four grid points:
\begin{equation*}
v^{n}_{\iph,j}=\frac{v_{i,\jph}^n+v_{i+1,\jmh}^n+v_{i,\jmh}^n+v_{i+1,\jmh}^n}{4}.
\end{equation*}
$u^n_d$, which is an approximation of $u^n$ at $\xy{d}{d}$, is calculated using the quadratic interpolation defined in \cite{min2007second}. The approximation of $v^n_d$ does not differ much from that of $u^n_d$.

\subsubsection{Construction of linear system before iterative procedure}
Based on the notation used in \ref{section:NumericMethods_XGFM_STOKES}, the following values are assigned: \begin{equation*}
a=[p],\quad \mathbf{c}=\left[\mu\nabla u\right],\quad \mathbf{d}=\left[ \mu\nabla v\right].
\end{equation*}
Furthermore, linear operator $\mathbf{A}$ and external force with correction term $b(a,\mathbf{c},\mathbf{d},f)$ are defined to be the same as those described in \ref{section:NumericMethods_XGFM_STOKES}. Under the assumption that the jump condition $a^{n+1},\mathbf{c}^{n+1},\mathbf{d}^{n+1}$ is known, \eqref{eq:discretize_NS_overview} can be discretized as
\begin{equation}
\begin{pmatrix}
\rho I   & 0 & 0 \\
0 &   \rho I & 0\\
0 &   0 & 0
\end{pmatrix}
\Uvect{n+1} -\begin{pmatrix}
\rho u_d^n\\\rho v_d^n\\0
\end{pmatrix}= \Delta t \left(\mathbf{A}\Uvect{n+1}  -b(a^{n+1},\mathbf{c}^{n+1},\mathbf{d}^{n+1}, -\rho \mathbf{g})\right).
\end{equation}
The saddle-point system is obtained when this discretization is organized as follows:
\begin{equation} \label{eq:NS_saddle_pt}
\begin{pmatrix}
\rho I -\Delta t (\mu \lap)^\text{GFM}  & 0 & \Delta t\nabla_x^h \\
0 &   \rho I -\Delta t (\mu \lap)^\text{GFM}  & \Delta t\nabla_y^h \\
-  \Delta t \nabla_x^h &  - \Delta t \nabla_y^h  & 0
\end{pmatrix}
\begin{pmatrix}
u^{n+1}\\v^{n+1}\\p^{n+1}
\end{pmatrix}=\begin{pmatrix}
\rho u_d^n\\\rho v_d^n\\0
\end{pmatrix} -\Delta t  b(a^{n+1},\mathbf{c}^{n+1},\mathbf{d}^{n+1},- \rho \mathbf{g}).
\end{equation}
One natural way of setting $\rho$ is to determine its value based on whether the grid points belong to either $\Omega^+$ or $\Omega^-$:
\begin{equation*}
\rho_{\iph,j}=\begin{cases}
\rho^- \text{ if } \phi_{\iph,j}^{n+1} <0\\ \rho^+ \text{ if } \phi_{\iph,j}^{n+1}\geq 0
\end{cases}.
\end{equation*}
However, our choice of $\rho$ near the interface differs from the $\rho$ obtained via the aforementioned method. First, the local truncation error of $p_x$ near the interface is considered. The following assumptions are then made: \(\phi_{i,j}^{n+1}<0,\phi_{i+1,j}^{n+1}>0\), and \(\phi_{\iph,j}^{n+1}<0\). Let
\[\theta = \frac{|\phi_{i,j}^{n+1}|}{|\phi_{i,j}^{n+1}|+|\phi_{i+1,j}^{n+1}| },\]
\begin{figure}
	\centering
	\includegraphics[width=0.5\textwidth]{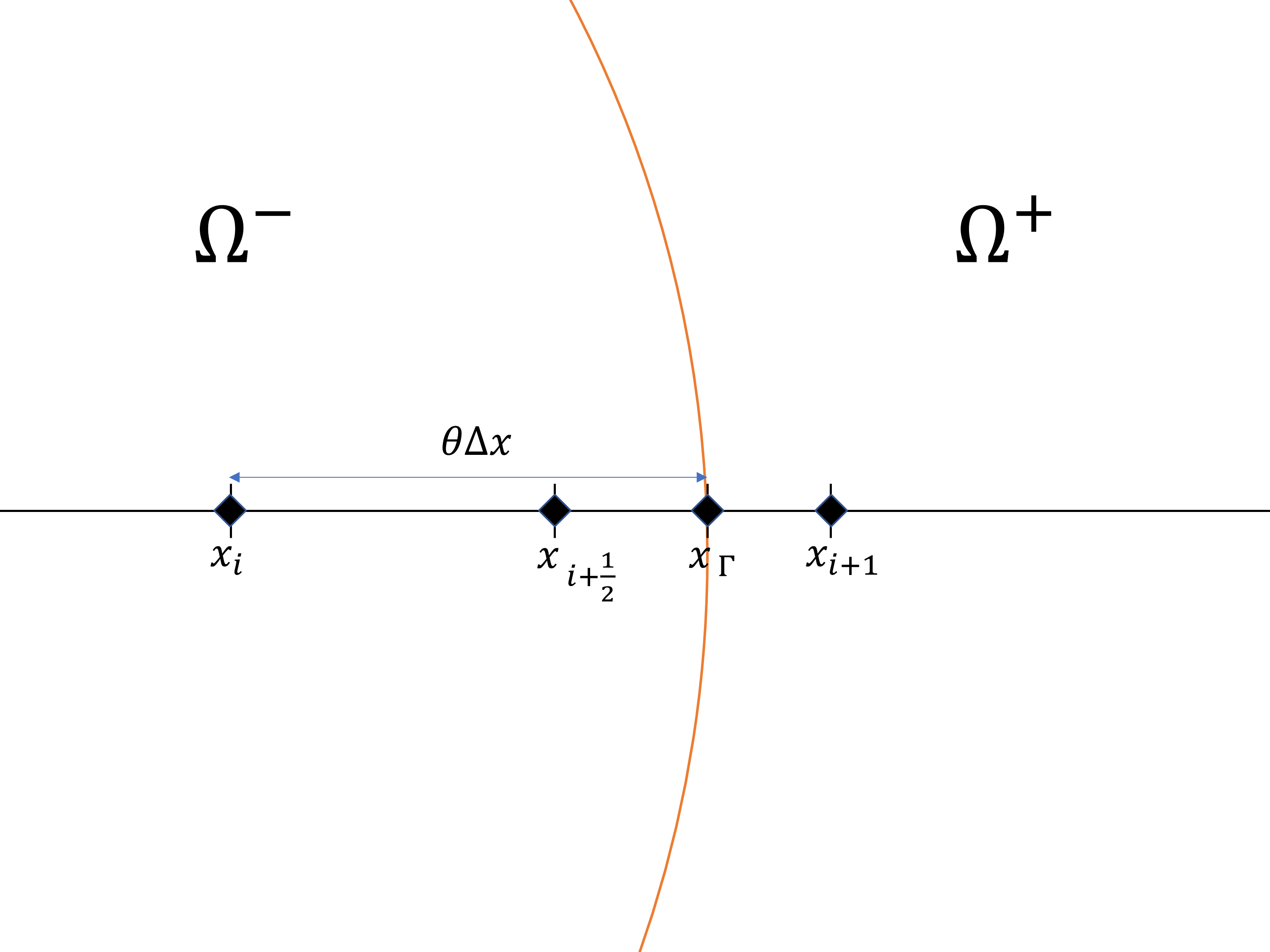}
	\caption{Visualization of the grid points near the interface when approximating $p_x^-$.}
	\label{fig:xgfmgrid}
\end{figure}
such that $x_\Gamma=(x_i+\theta \Delta x,y_j)$ approximates the location of the interface. Figure \ref{fig:xgfmgrid} shows a visualization of these grid points and interface. $p_x^- $ at $\xy{\iph}{j}$ is approximated as
\[p_x^- \approx \frac{p_{i+1,j}-\left[p\right]_\Gamma-p_{i,j}}{\Delta x}.\]
The value $p_R^\pm=p^\pm\left(x_i+\theta \Delta x,y_j\right)$ is then assigned. Thus,
\begin{align*}
\frac{p_{i+1,j}-\left[p\right]_\Gamma-p_{i,j}}{\Delta x}-p_x^- &= \frac{p_R^+ +(1-\theta)p_x^+ \Delta x+ -\left[p\right]_\Gamma-p_{i,j}}{\Delta x}-p_x^- +O(\Delta x)\\
&=\frac{p_R^--p_{i,j}}{\Delta x} +(1-\theta)p_x^+-p_x^- +O(\Delta x)\\
& = \theta\frac{p_R^--p_{i,j}}{\theta\Delta x} +(1-\theta)p_x^+ -p_x^-+O(\Delta x) \\
&= \theta p_x^- +(1-\theta)p_x^+ -p_x^-+o(\Delta x)=(1-\theta)[p_x] +O(\Delta x).
\end{align*}
If the jump conditions on the Navier--Stokes equation are considered, 
\begin{equation*}
\left[\rho \frac{d\uvec}{dt}\right] =\left[\mu\lap \uvec -\nabla p -\rho \mathbf{g}\right]
\end{equation*}
is determined. Afterward, when the material derivative being continuous across the interface \(\left[\frac{d\uvec}{dt}\right]=0\) is considered, 
\begin{equation*}
\left[\nabla p\right]=-\left[\rho \right]\left(\frac{d\uvec}{dt}+ \mathbf{g}\right) +\left[\mu\lap \uvec  \right]
\end{equation*}
is obtained. In real-world simulations, \(\left[\rho\right]\) dominates \([\mu \lap \uvec]\), and thus $\rho^{n+1}_{\iph,j}$ is selected to reduce the truncation error of $p_x$:
\begin{equation*}
\rho^{n+1}_{\iph,j}=\theta\rho^- + (1-\theta)\rho^+.
\end{equation*}
If the truncation error of the material derivative at $\xy{\iph}{j}$ is then computed, 
\begin{equation*}
\begin{aligned}
\rho_{\iph,j}^{n+1}\left(\frac{u_{\iph,j}^{n+1}-u^n_d}{\Delta t}\right)-\rho^- \frac{du}{dt}&=(\rho_{\iph,j}^{n+1}-\rho^-) \frac{du}{dt}+ O(\Delta x + \Delta y)\\
&=(1-\theta)[\rho]\frac{du}{dt}+ O(\Delta x + \Delta y)
\end{aligned}
\end{equation*}
is obtained, which cancels the local truncation error of $p_x$ with the $[\rho]$ term. Generally, $\rho_{\iph,j}^{n+1}$ is defined as 
\begin{equation*}
\rho_{\iph,j}^{n+1}=\theta \rho^1 +(1-\theta)\rho^2
\end{equation*}
for
\begin{equation*}
\rho^1=\begin{cases}
\rho^-\text{ if }\phi^{n+1}_{i,j}<0  \\\rho^+ \text{ if }\phi^{n+1}_{i,j}\geq0
\end{cases},\quad
\rho^2=\begin{cases}
\rho^-\text{ if }\phi^{n+1}_{i+1,j}<0\\ \rho^+ \text{ if }\phi^{n+1}_{i+1,j}\geq0
\end{cases}
\end{equation*}
and
\[\theta = \frac{|\phi_{i,j}^{n+1}|}{|\phi_{i,j}^{n+1}|+|\phi_{i+1,j}^{n+1}| }.\]
$\rho_{i,\jph}^{n+1}$ is defined similarly with few modifications.

\subsubsection{Iterative method}
Given that linear system \eqref{eq:NS_saddle_pt} has been established, the iterative method from \ref{subsec:Stokes_iterative} can be applied. Let 
\begin{equation*} 
\hat{\mathbf{A}}\Uvec{}=
\begin{pmatrix}
\rho I -\Delta t (\mu \lap)^\text{GFM}  & 0 & \Delta t\nabla_x^h \\
0 &   \rho I -\Delta t (\mu \lap)^\text{GFM}  & \Delta t\nabla_y^h \\
-  \Delta t \nabla_x^h &  - \Delta t \nabla_y^h  & 0
\end{pmatrix}\Uvect{}
\end{equation*}
and
\[ \hat{b}(a,\mathbf{c},\mathbf{d},\rho \mathbf{g})=\begin{pmatrix}
\rho u_d^n\\\rho v_d^n\\0
\end{pmatrix} -\Delta t  b(a,\mathbf{c},\mathbf{d}, -\rho \mathbf{g}).\]
To sum up, the velocity and pressure at time-level $t^{n+1}$ are computed via the solution for $u^{n+1},v^{n+1},p^{n+1}$ of the linear system 
\begin{equation}\label{eq:iterative_system}
\hat{\mathbf{A}} \Uvec{n+1}=\hat{b}(a^{n+1},\mathbf{c}^{n+1},\mathbf{d}^{n+1},-\rho \mathbf{g}),
\end{equation}
where
\begin{equation}\label{eq:iterative_jmp_cond}
\begin{gathered}
\mathbf{c}^{n+1}=\begin{pmatrix}
[\mu](2n_x^2 n_y -n_y) u_\tanv^{\text{ext}} & + [\mu](n_x n_y^2 -n_x^3) v_\tanv^{\text{ext}} \\
[\mu](2n_x n_y^2 + n_x) u_\tanv^{\text{ext}} & + [\mu]( n_y^3 -n_x^2n_y) v_\tanv^{\text{ext}} 
\end{pmatrix} \\
\mathbf{d}^{n+1} =\begin{pmatrix}
[\mu](-2n_x^2 n_y -n_y) v_\tanv^{\text{ext}}&  + [\mu](n_x n_y^2 -n_x^3) u_\tanv^{\text{ext}} \\
[\mu](-2n_x n_y^2 + n_x) v_\tanv^{\text{ext}}&  + [\mu]( n_y^3 -n_x^2 n_y) u_\tanv^{\text{ext}} 
\end{pmatrix} \\
a^{n+1}= -2[\mu](-n_y u_\tanv^{\text{ext}} +n_x v_\tanv^{\text{ext}})-\beta \kappa  .
\end{gathered}
\end{equation}
$u^\text{ext}_\tanv,v^\text{ext}_\tanv$ are computed according to \eqref{eq:extrapolation_boundary} with the steady-state solution of \eqref{eq:psuedo_extra}, where the boundary condition is given by $u_\Gamma=u^{n+1},v_\Gamma=v^{n+1}$. Because the right-hand side of \eqref{eq:iterative_system} also involves a linear combination of $u^{n+1},v^{n+1}$, an iterative method is used to solve the given linear system. For the iterative method, the time-level subscript $n+1$ is dropped for the sake of simplicity. $k$ denotes the number of iterative procedures.
\begin{enumerate}
	\item For $a^{k},\mathbf{c}^{k},$ and $\mathbf{d}^{k}$, solve for $\tilde{u}^{k+1},\tilde{v}^{k+1},\tilde{p}^{k+1}$ in
	\begin{equation*}
	\hat{\mathbf{A}} \left(\tilde{u}^{k+1} ,\tilde{v}^{k+1}, \tilde{p}^{k+1}\right)=\hat{b}(a^{k},\mathbf{c}^{k},\mathbf{d}^{k},-\rho \mathbf{g}).
	\end{equation*}
	At the beginning of the iterative method, set $\mathbf{c}^0=\mathbf{0},\mathbf{d}^0=\mathbf{0}$, and $a^0=-\beta \kappa$.
	\item Get $\tilde{u}^\text{ext},\tilde{v}^\text{ext}$ as a steady-state solution of velocity extrapolation \ref{subsec:velocity_extrapolation}, with boundary condition $u_\Gamma =\tilde{u}^{k+1},v_\Gamma=\tilde{v}^{k+1}$. Compute the jump condition \(\tilde{a}^{k+1},\tilde{\mathbf{c}}^{k+1},\tilde{\mathbf{d}}^{k+1}\) using \eqref{eq:reformulation_jmp}, where $u_\tau, v_\tau$ are substituted with $\tilde{u}^\text{ext},\tilde{v}^\text{ext}$.	
	\item For the two residual vectors of \eqref{eq:iterative_system}, 
	\begin{equation*}
	\begin{aligned}
	r^{k}&=\hat{b}(a^{k},\mathbf{c}^{k},\mathbf{d}^{k},-\rho \mathbf{g})-\hat{\mathbf{A}} \Uvec{k},\\
	\tilde{r}^{k+1}&=\hat{b}(\tilde{a}^{k+1},\tilde{\mathbf{c}}^{k+1},\tilde{\mathbf{d}}^{k},-\rho \mathbf{g})-\hat{\mathbf{A}} \left(\tilde{u}^{k+1}, \tilde{v}^{k+1}, \tilde{p}^{k+1}\right),
	\end{aligned}
	\end{equation*}
	and with respect to $\left(u^k,v^k,p^k\right)$ and $\left(\tilde{u}^{k+1},\tilde{v}^{k+1},\tilde{p}^{k+1}\right)$,
	find $\eta_k$ that minimizes $\norm{(1-\eta_k)r^k + \eta_k \tilde{r}^{k+1}}_2$, which is given by \eqref{eq:eta_formula}. The $(k+1)$-th iterative solutions $u^{k+1},v^{k+1},p^{k+1}$, and jump conditions $a^{k+1},\mathbf{c}^{k+1},\mathbf{d}^{k+1}$ are computed as described in \ref{subsec:Stokes_iterative}, as a linear combination with weights $1-\eta_k$ and $\eta_k$. 
\end{enumerate}
The aforementioned three steps are repeated until convergence occurs. Because $\hat{\mathbf{A}}$ is symmetric, we use the minimal residual (MINRES) method to solve the linear system in step 1 of the iterative method. An initial guess for the iterative method is given by the velocity and pressure at time level $t^{n}$. As a by-product of the iterative method, $u^\text{ext},v^\text{ext}$ at time level $t^{n+1}$ is obtained. These values are used to advect the level-set from $t^{n+1}$ to $t^{n+2}$, via the definitions 
\( u_{i,j}^{n+1} = \frac{u^{\text{ext}}_{\iph,j}+u^{\text{ext}}_{\imh,j}}{2},\quad  v_{i,j}^{n+1} = \frac{v^{\text{ext}}_{i,\jph}+v^{\text{ext}}_{i,\jmh}}{2}\). 

\subsubsection{Time-step restriction}
In \cite{kang2000boundary}, time-step restrictions were computed for each convection, viscous term, gravity, and surface tension. The classical time-step restriction of \cite{brackbill1992continuum} was not violated, but the effects of external forces and convection were summed up to produce a small time step $\Delta t$. On the other hand, our time-step restriction is based on \cite{theillard2019sharp}, which extended the time-step restriction created by \cite{galusinski2008stability}, which, in turn, is valid for two-phase flows involving the same densities and viscosities.
\[C_{cfl}=c_0 \frac{1}{\norm{\uvec}_\infty}\Delta x\]
and
\[S_{cfl}=\frac{c_1\mu_{\min}}{\beta}\Delta x + \sqrt{\left(c_1\frac{\mu_{\min}}{\beta}\Delta x\right)^2+c_2\frac{\left(\rho^-+\rho^+\right) \Delta x^3}{4 \pi\beta} }\]
are then defined as the time-step restrictions for convection and surface tension, respectively. The researchers in \cite{theillard2019sharp} proposed to set time-step 
\[\Delta t =\min(C_{cfl},S_{cfl})\]
with $c_1,c_2<1$ and $c_0<1$. Because our method advects the level-set with the Eulerian method, $c_0<0.5$ must be imposed. We use $c_0=0.45, c_1=0.9$, and $c_2=0.9$ in our numerical simulations.

\section{Numerical Experiments}\label{sec:Numerical experiment}
In this section, we present our numerical experiments on two-phase incompressible flows. First, we deal with analytical solutions, and then singular source and external force terms are given according to the solutions. Practical numerical examples with surface tension and gravity are then considered. All computations are performed in C++ and implemented on a personal computer with 3.40 GHz CPU and 32.0 GB memory without parallelization.
\subsection{Steady-state Stokes equation}\label{ex:steady_state_stoke}
\begin{figure}
	\centering
	\subfloat[a][$u$]{	\includegraphics[width=0.3\textwidth]{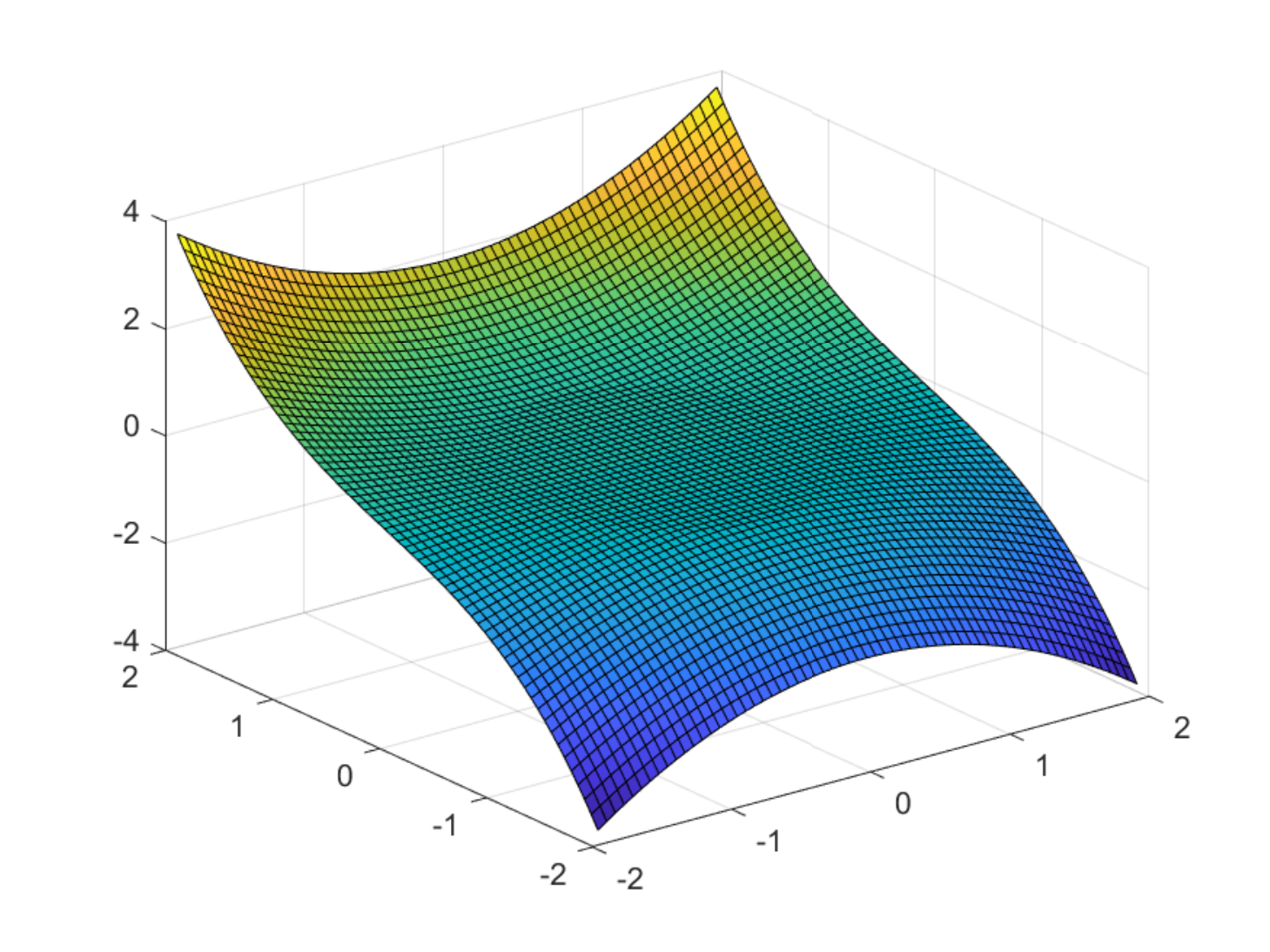}}
	\subfloat[b][$v$]{\includegraphics[width=0.3\textwidth]{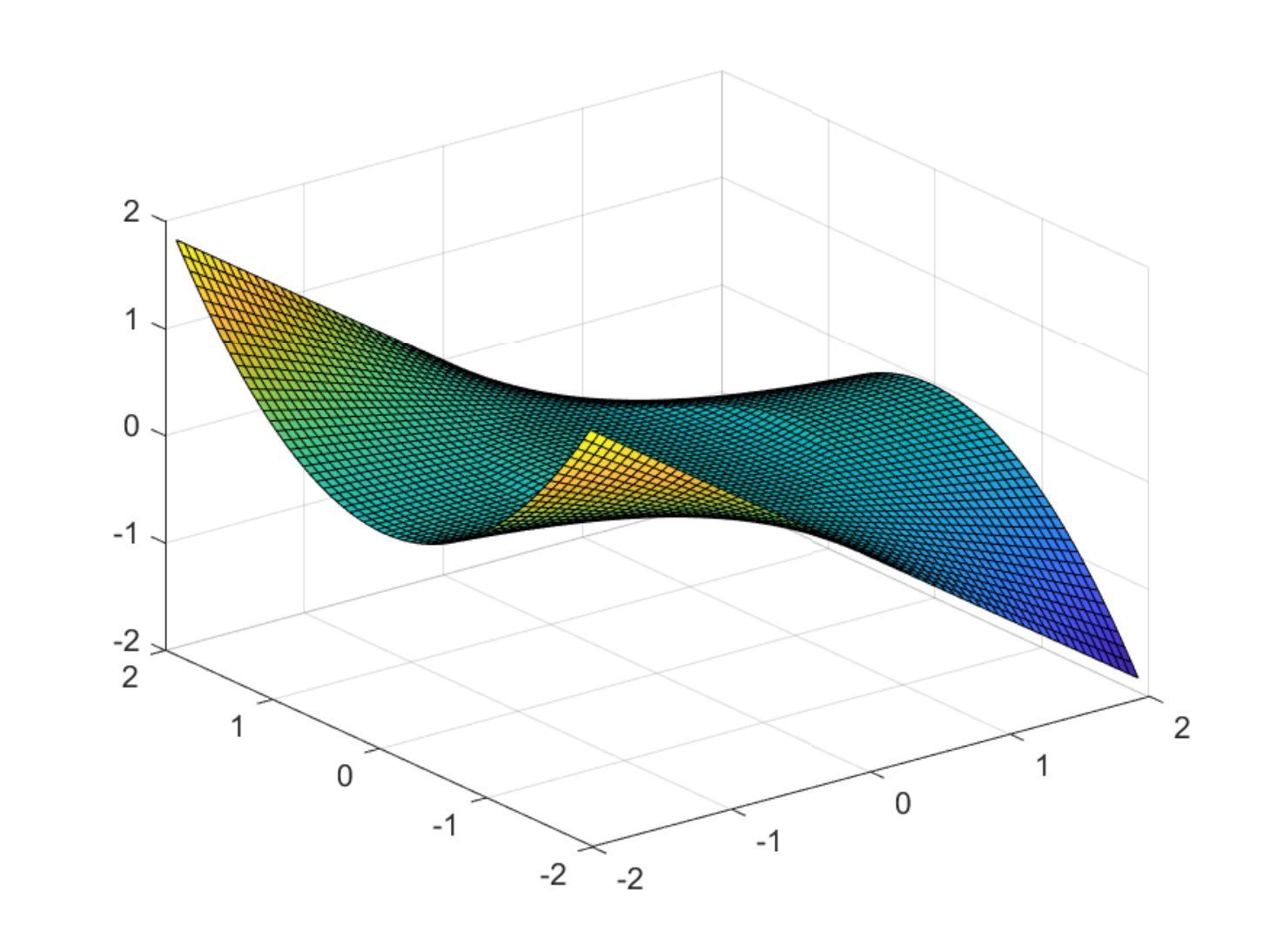}}
	\subfloat[c][$p$]{\includegraphics[width=0.3\textwidth]{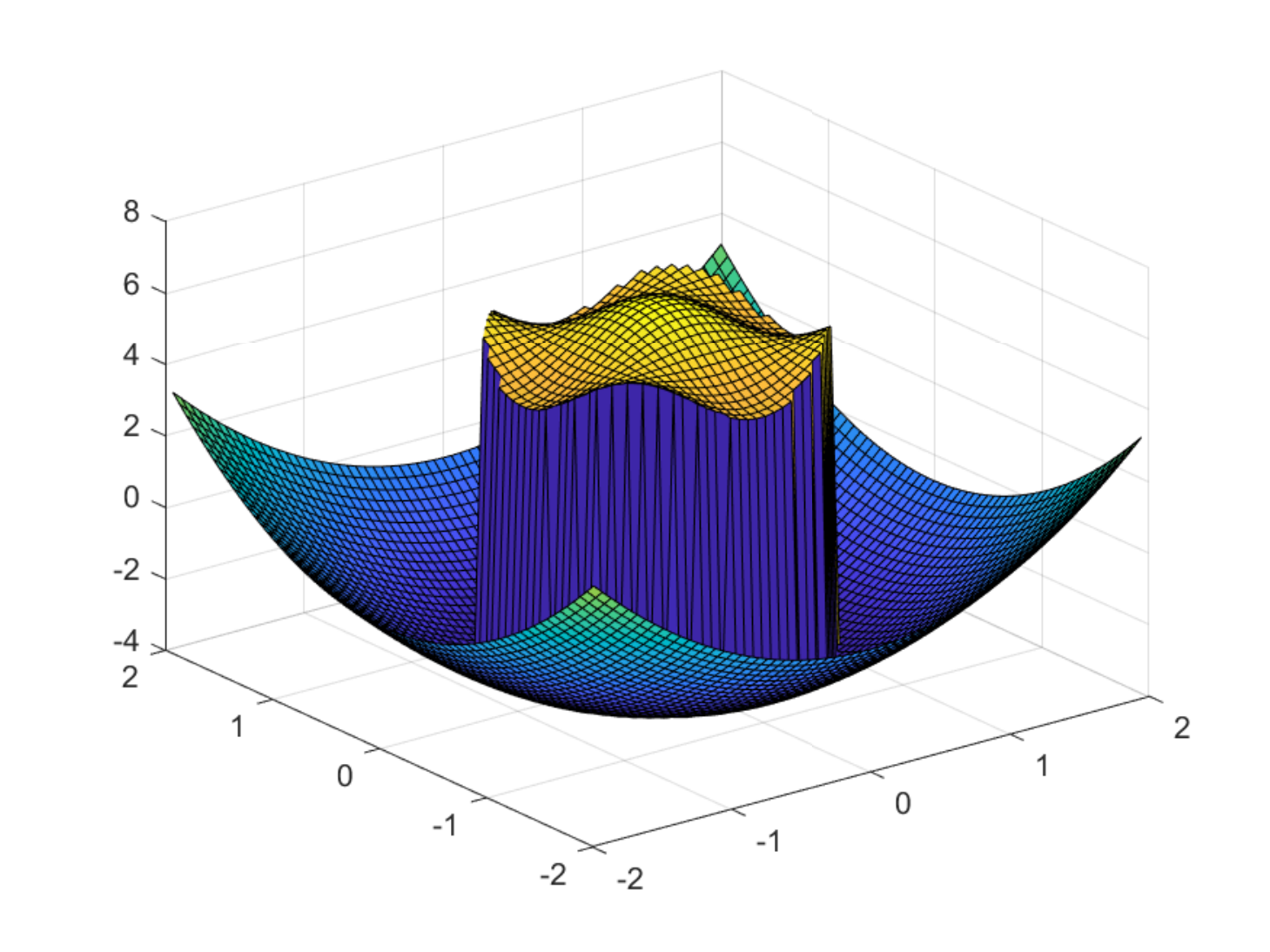}}
	\caption{Visualization of solution $\uvec=(u,v)$ and $p$ for \ref{ex:steady_state_stoke}.}
	\label{fig:stokes_sol}
\end{figure}

We first consider the two-phase steady-state Stokes equation \(\mu \lap \uvec - \nabla p =\mathbf{f}\) on \(\Omega= \left[-2,2\right]^2\). The interface $\Gamma$ is defined by the zero level-set of $\phi=\sqrt{x^2+y^2}-1$, and the exact solution $\uvec=\left(u,v\right)$ and $p$ are
\begin{equation*}
\begin{gathered}
u=\begin{cases}
\frac{y}{4} &\text{ if } \phi(x,y)<0 \\
\frac{y}{4}\left(x^2+y^2\right) &\text{ if } \phi(x,y)\geq0 
\end{cases},\quad
v=\begin{cases}
-\frac{x}{4}(1-x^2) &\text{ if } \phi(x,y)<0 \\
-\frac{x}{4}y^2 &\text{ if } \phi(x,y)\geq0 
\end{cases}, \quad
p=\begin{cases}
\cos(\pi x) \cos(\pi y)+10&\text{ if } \phi(x,y)<0 \\
x^2+y^2 &\text{ if } \phi(x,y)\geq0 
\end{cases}.
\end{gathered}
\end{equation*}
Figure \ref{fig:stokes_sol} shows the solution profile on a $64\times 64$ grid. The external force term $\mathbf{f}=(f_1,f_2)$ and the jump condition $\left[\mu  \left(\nabla\uvec +\nabla\uvec^T\right) \normal - p \normal \right]=\mathbf{G}=(G_1, G_2)$ are obtained according to the exact solution:
\begin{equation*}
\begin{gathered}
f_1=\begin{cases}
\pi\sin(\pi x) \cos(\pi y) &\text{ if } \phi(x,y)<0 \\
2\mu^+ y -2x&\text{ if } \phi(x,y)\geq0 
\end{cases},\quad
f_2=\begin{cases}
\frac{3\mu^- x}{2} +\pi\cos(\pi x) \sin(\pi y) &\text{ if } \phi(x,y)<0 \\
-\frac{\mu^+ x}{2} -2y &\text{ if } \phi(x,y)\geq0 
\end{cases}\\
G_1= x\,\left(\cos(\pi x) \cos(\pi y)-x^2-y^2+10+\mu ^+\,x\,y\right)+y\,\left(\mu^+\,\left(\frac{x^2}{4}+\frac{y^2}{2}\right)-\frac{3\,\mu^- \,x^2}{4}\right),\\
G_2=    y\,\left(\cos(\pi x) \cos(\pi y)-x^2-y^2+10-\mu ^+\,x\,y\right)+x\,\left(\mu ^+\,\left(\frac{x^2}{4}+\frac{y^2}{2}\right)-\frac{3\,\mu ^-\,x^2}{4}\right).
\end{gathered}
\end{equation*}
Numerical experiments are conducted with viscosities $\mu^+=0.1,\mu^-=0.01$ or $\mu^+=0.01,\mu^-=0.1$. The results of convergence for the $L^\infty$ and $L^2$ norms are presented in Table \ref{tab:stokes}. The estimated order of convergence for pressure in the $L^\infty$ norms is around $0.8$, but first-order convergence is observed for pressure in the $L^2$ norms and for velocities in both the $L^\infty$ and $L^2$ norms. In Table \ref{tab:stokes_precond}, we present the cumulative numbers of MINRES iterations when preconditioners are chosen as incomplete Cholesky decompositions of $M$, with $\alpha= \frac{1}{\Delta x^2}$ defined as in \eqref{eq:preconditioner}. We also observe that the growth rate of the total iteration is linear to the grid size.
\begin{table}[]
	\caption{Convergence of Example \ref{ex:steady_state_stoke}}
	\label{tab:stokes}
	\begin{center}
		\subfloat[][$\mu^+=0.1,\mu^-=0.01$]{
			
			\begin{tabular}{lllllllll}
				\hline
				
				resolution &   $L^\infty$ error of $\uvec$  & order     &   $L^\infty$ error of $p$        & order     &  $L^2$ error of $\uvec$        &  order    & $L^2$ error of $p$        &  order   \\
				\noalign{\smallskip}\hline\noalign{\smallskip}
				$32^2$                 & 9.56.E-02 &      & 9.19.E-02 &      & 1.83.E-01 &      & 8.08.E-02 &      \\
				$64^2$                 & 4.67.E-02 & 1.03 & 5.42.E-02 & 0.76 & 7.04.E-02 & 1.37 & 3.12.E-02 & 1.37 \\
				$128^2$                & 1.68.E-02 & 1.48 & 2.41.E-02 & 1.17 & 2.03.E-02 & 1.79 & 1.02.E-02 & 1.61 \\
				$256^2$                & 7.46.E-03 & 1.17 & 1.59.E-02 & 0.59 & 9.34.E-03 & 1.12 & 4.55.E-03 & 1.17
			\end{tabular}
		}
		
		\subfloat[][$\mu^+=0.01,\mu^-=0.1$]{
			\begin{tabular}{lllllllll}
				\hline
				resolution &   $L^\infty$ error of $\uvec$  & order     &   $L^\infty$ error of $p$        & order     &  $L^2$ error of $\uvec$        &  order    & $L^2$ error of $p$        &  order   \\
				\noalign{\smallskip}\hline\noalign{\smallskip}
				$32^2$                 & 1.35.E-01 &      & 7.61.E-02 &      & 2.78.E-01 &      & 5.48.E-02 &      \\
				$64^2$                 & 5.72.E-02 & 1.24 & 4.06.E-02 & 0.91 & 9.82.E-02 & 1.50 & 1.59.E-02 & 1.78 \\
				$128^2$                & 1.95.E-02 & 1.55 & 2.49.E-02 & 0.70 & 2.94.E-02 & 1.74 & 9.35.E-03 & 0.77 \\
				$256^2$                & 8.38.E-03 & 1.22 & 1.43.E-02 & 0.81 & 1.40.E-02 & 1.07 & 3.83.E-03 & 1.29
			\end{tabular}
		}
	\end{center}
\end{table}

\begin{table}[]
	\caption{Cumulative number of MINRES iterations}
	\label{tab:stokes_precond}
	\begin{center}
		\begin{tabular}{|c|c|c|}
			\hline
			
			resolution &  $\mu^+=0.1,\mu^-=0.01$   & $\mu^+=0.01,\mu^-=0.1$     \\
			\hline
			$32^2$                 & 1419  &  1495  \\ \hline
			$64^2$                 & 2638  & 3191 \\ \hline
			$128^2$                & 5084 & 6299 \\ \hline
			$256^2$                & 10038 & 10670 \\ \hline
			
		\end{tabular}
	\end{center}
\end{table}

\subsection{Analytical solution of Navier--Stokes equation with stationary interface}\label{ex:NS_analytic}
\begin{figure}
	\centering
	\subfloat[a][$u$]{	\includegraphics[width=0.3\textwidth]{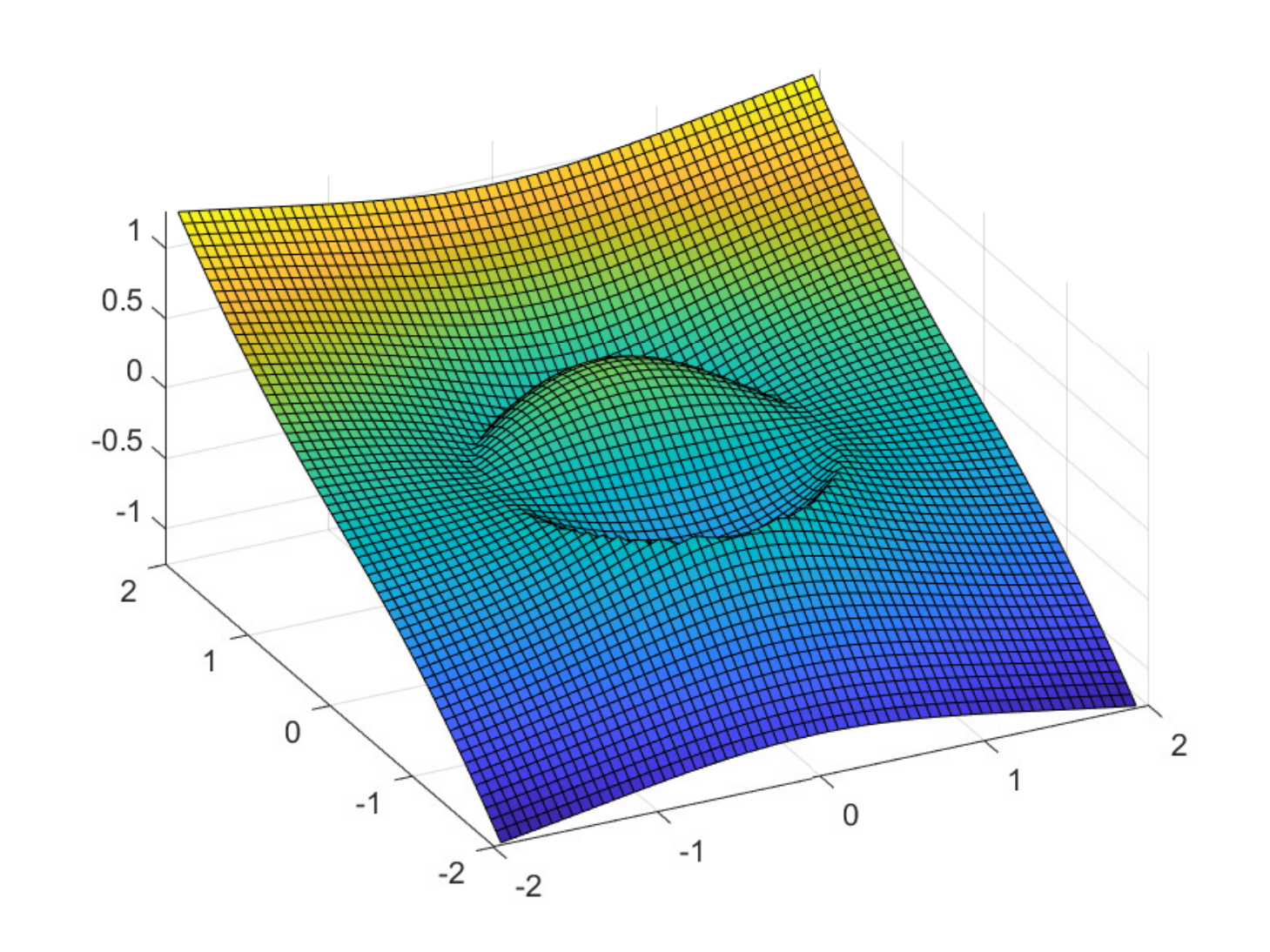}}
	\subfloat[b][$v$]{\includegraphics[width=0.3\textwidth]{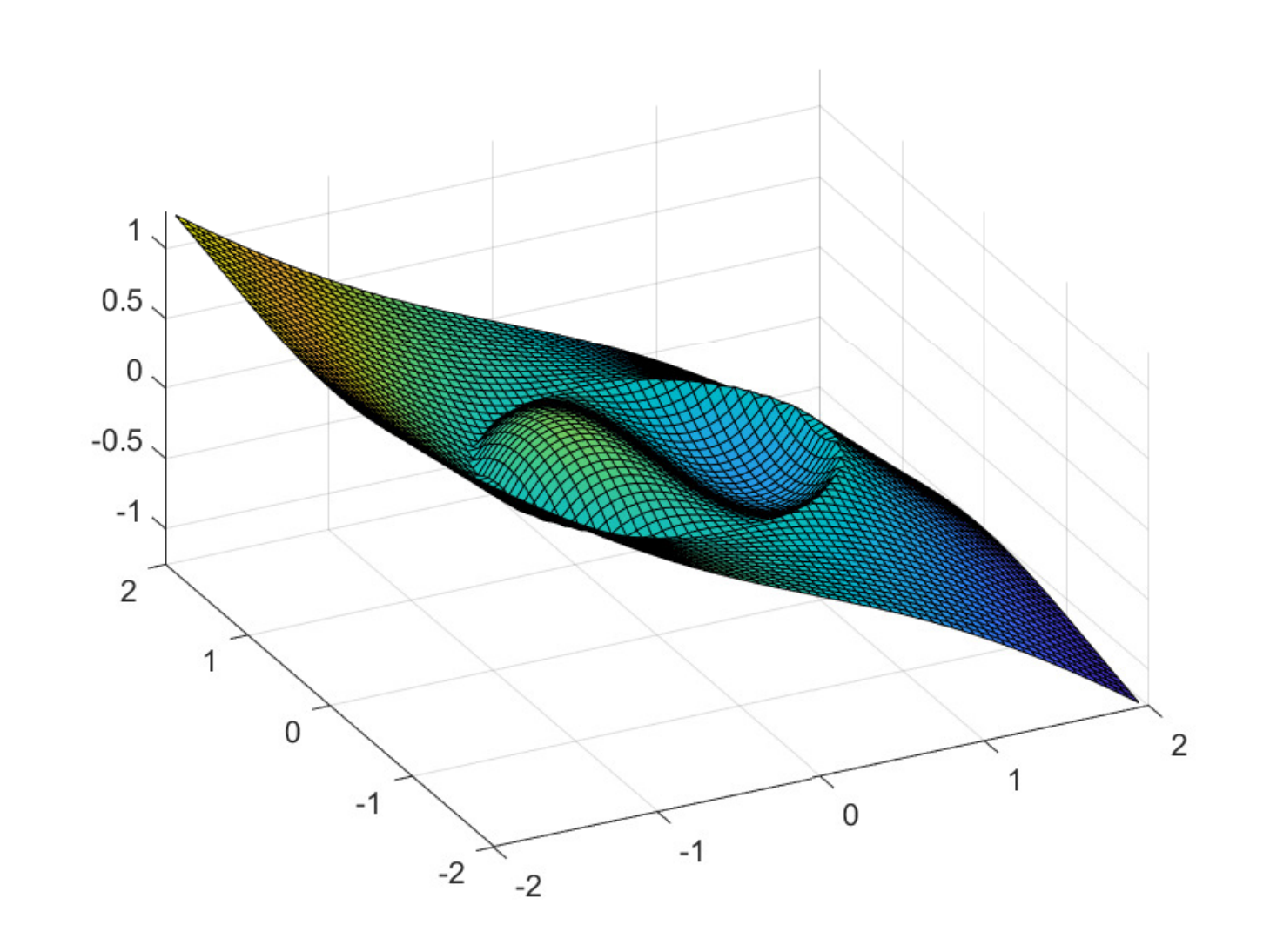}}
	\subfloat[c][$p$]{\includegraphics[width=0.3\textwidth]{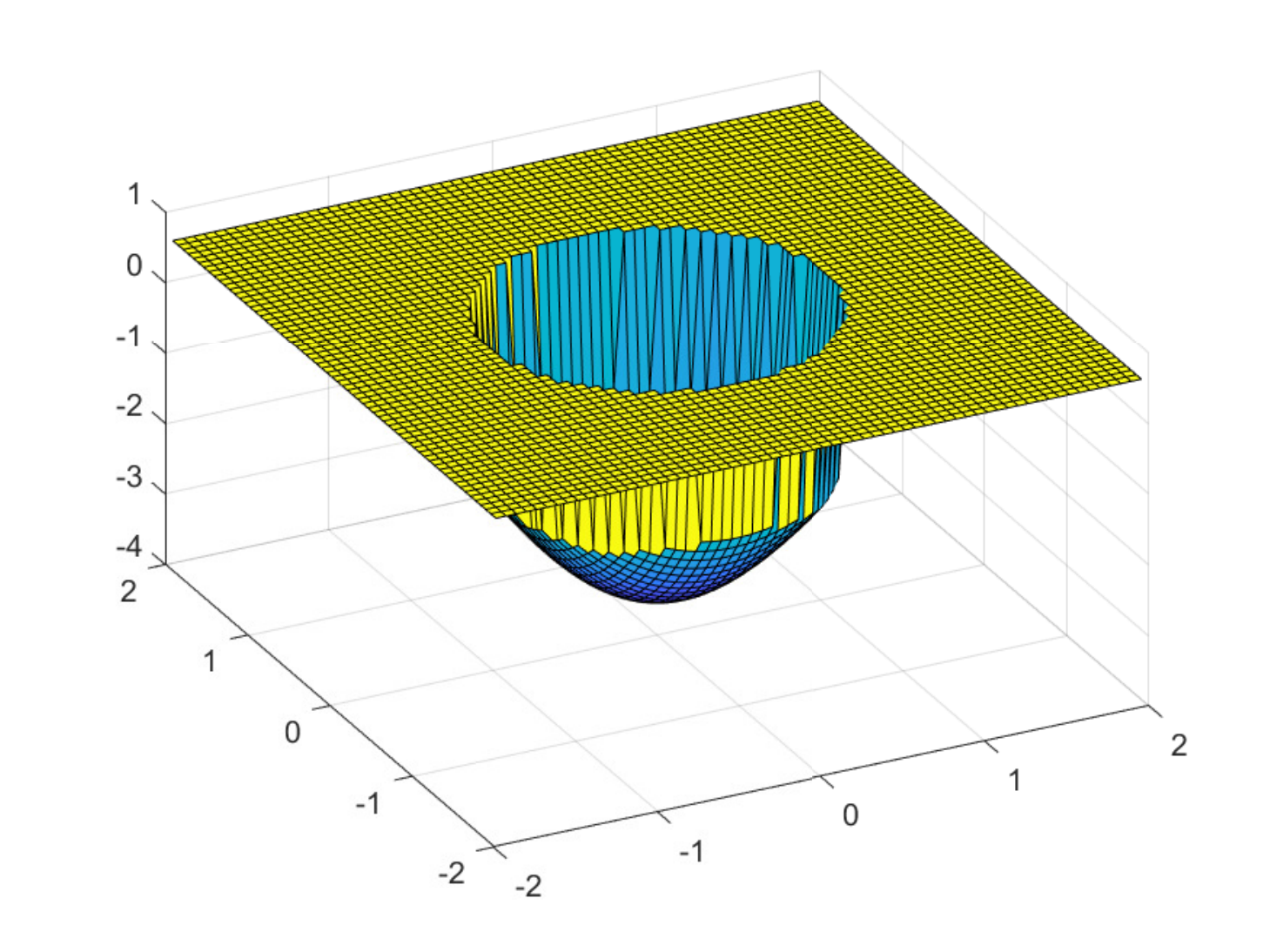}}
	\caption{Visualization of solution for Example \ref{ex:NS_analytic} at $t=\pi$.}
	\label{fig:NS_sol}
\end{figure}

As a second example, a two-phase incompressible Navier--Stokes equation with stationary interface $\Gamma=\{\left(x,y\right)\in \mathbb{R}^2 | \sqrt{x^2+y^2}=1\}$ is considered on the computational domain $\Omega=[-2,2]^2$. The exact solutions for velocity and pressure are determined to be 
\begin{equation*}
\begin{gathered}
u=\begin{cases}
y (x^2 + y^2- 1)\cos(t) &\text{ if } \phi(x,y)<0 \\
\left(\frac{y}{r}-y\right) \cos(t)&\text{ if } \phi(x,y)\geq0 
\end{cases},\quad
v=\begin{cases}
-x (x^2 + y^2- 1)\cos(t) &\text{ if } \phi(x,y)<0 \\
\left(-\frac{x}{r}+x\right) \cos(t)&\text{ if } \phi(x,y)\geq0 
\end{cases},\\
p=\begin{cases}
\cos(x) \cos(y)\cos(t) &\text{ if } \phi(x,y)<0 \\
0 &\text{ if } \phi(x,y)\geq0 
\end{cases}
\end{gathered}
\end{equation*}
where $r=\sqrt{x^2+y^2}$.
The density and viscosity are chosen to be \(\frac{\rho^+}{\rho^-}=\frac{\mu^+}{\mu^-}=100\) for $\rho^+=1$ and $\mu^+ =0.01$. As in \ref{ex:steady_state_stoke}, the values of $\mathbf{f}$ and $\mathbf{G}$ are determined according to the analytical solution. Furthermore, $\uvec=\mathbf{0}$ on $\Gamma$, and therefore the interface does not change over time. To check the accuracies of solution methods for the Navier--Stokes equation, we do not advect level-set in this example. Numerical simulations are performed up to $t=\pi$, and the profile of the solution is shown in Figure \ref{fig:NS_sol}. First-order convergence for velocity and pressure both in $L^\infty$ and $L^2$ norms are observed in Table \ref{tab:ns_analytic1}. The results demonstrate that our method can manage the non-smoothness of solutions to two-phase Navier--Stokes equations.
\begin{table}[]
	\caption{Convergence of Example \ref{ex:NS_analytic}.}
	\label{tab:ns_analytic1}
	\begin{center}
		\begin{tabular}{lllllllll}
			\hline
			resolution &   $L^\infty$ error of $\uvec$  & order     &   $L^\infty$ error of $p$        & order     &  $L^2$ error of $\uvec$        &  order    & $L^2$ error of $p$        &  order   \\
			\noalign{\smallskip}\hline\noalign{\smallskip}
			$32^2$   & 1.54.E-01 & & 3.20.E-02 &  & 1.35.E-01 &           & 3.48.E-02 &                 \\
			$64^2$   & 8.62.E-02 & 0.84      & 1.85.E-02 & 0.79      & 7.20.E-02 & 0.91      & 1.35.E-02 & 1.36 \\
			$128^2$ & 3.62.E-02 & 1.25      & 9.47.E-03 & 0.97      & 2.91.E-02 & 1.31      & 6.53.E-03 & 1.05 \\
			$256^2$ & 1.75.E-02 & 1.05      & 4.93.E-03 & 0.94      & 1.35.E-02 & 1.11      & 3.11.E-03 & 1.07
		\end{tabular}
	\end{center}
\end{table}

\subsection{Analytical solution of Navier--Stokes equation with moving interface}\label{ex:NS_analytic2}
\begin{figure}
	\subfloat[a][$u$]{	\includegraphics[width=0.3\textwidth]{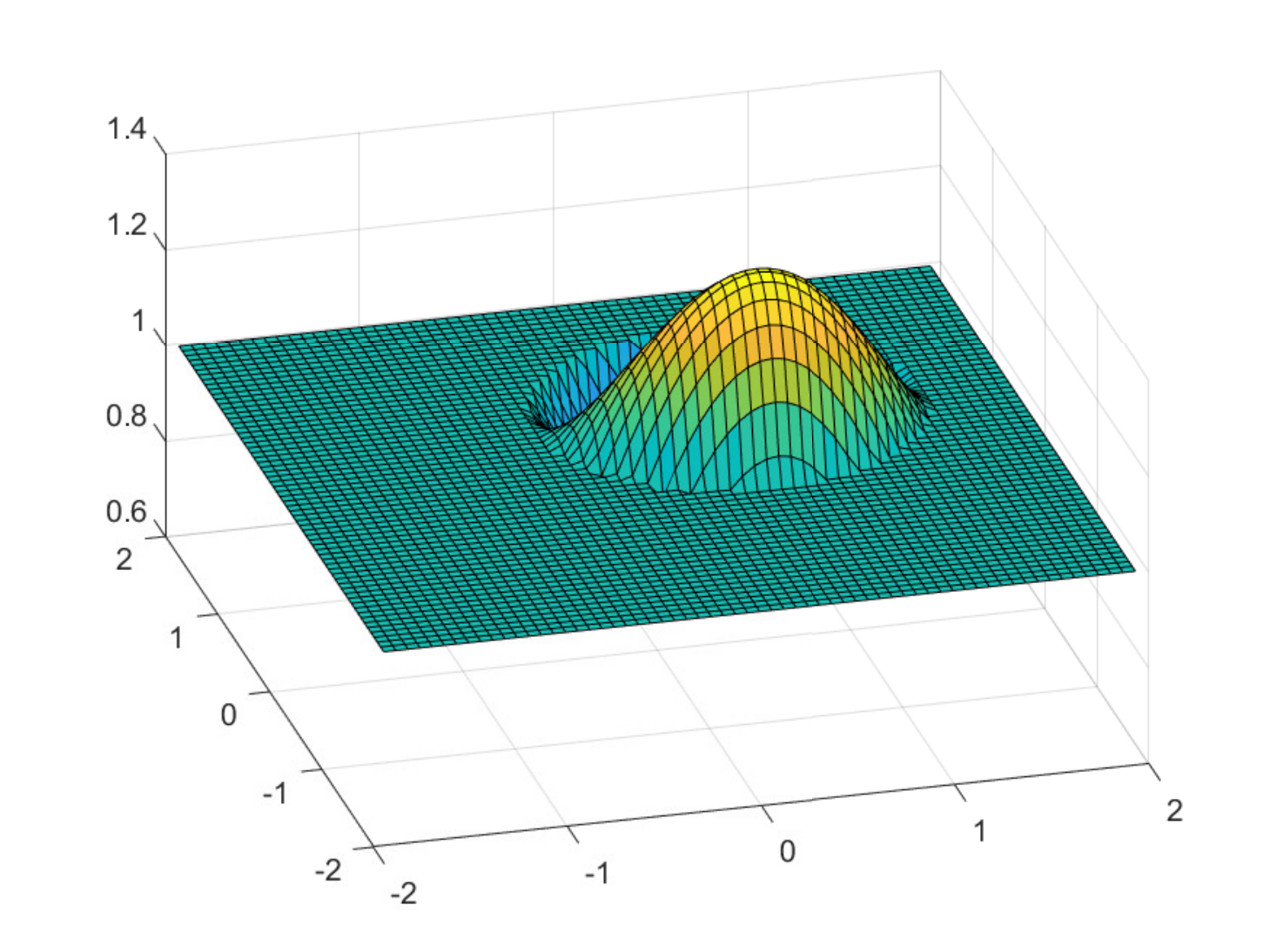}}
	\subfloat[b][$v$]{\includegraphics[width=0.3\textwidth]{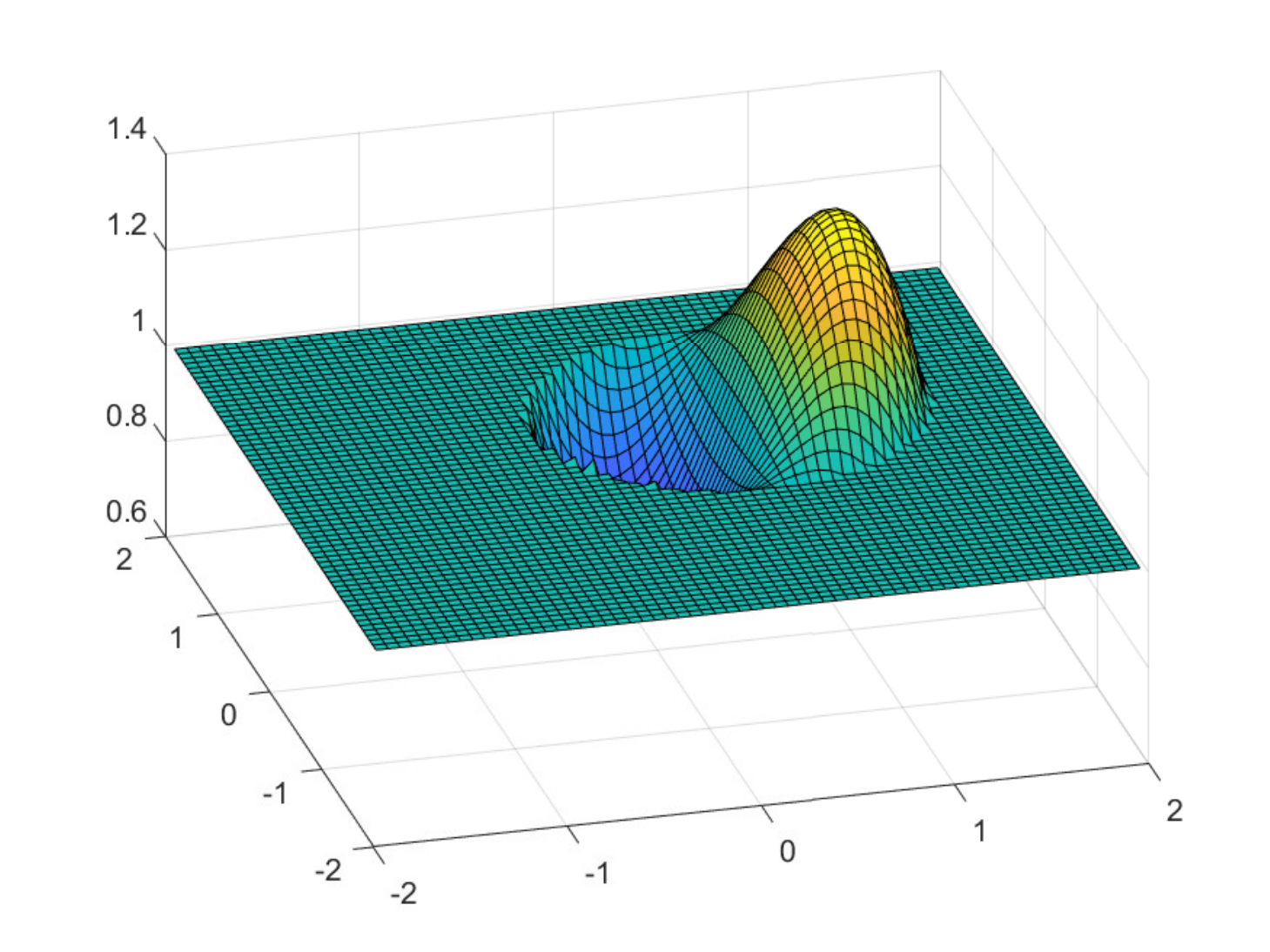}}
	\subfloat[c][$p$]{\includegraphics[width=0.3\textwidth]{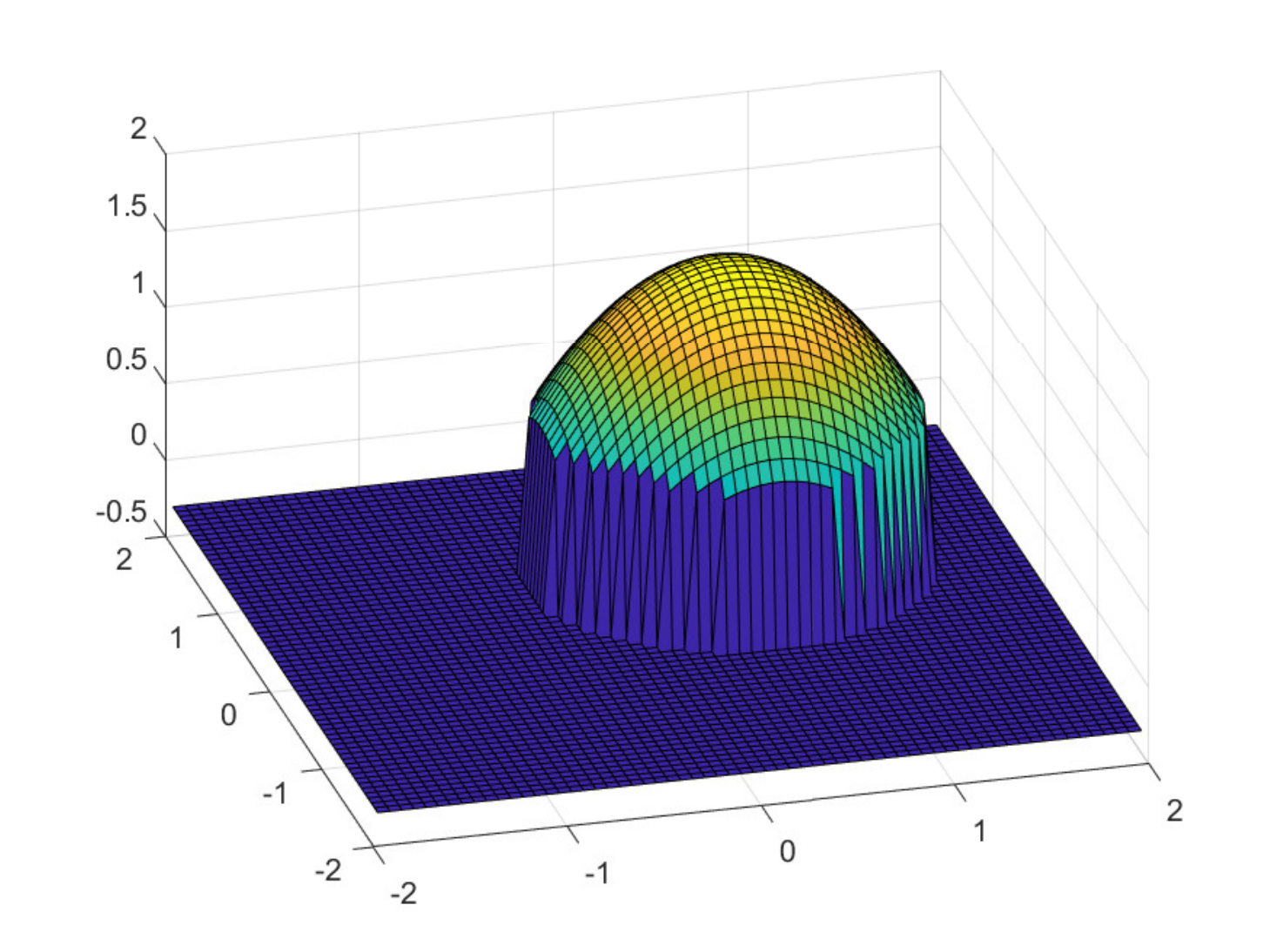}}
	\caption{Visualization of solution for Example \ref{ex:NS_analytic2} at $t=1$.}
	\label{fig:analytic_moving}
\end{figure}

We now consider moving interface $\Gamma_t=\{\left(x,y\right)\in \mathbb{R}^2 | \sqrt{(x-t+0.5)^2+(y-t+0.5)^2}=1\}$ on the computational domain $\Omega=[-2,2]^2$. The exact solutions are chosen to have a constant velocity on the interface
\begin{equation*}
\begin{gathered}
u=\begin{cases}
(y-t+0.5) (r^2-1)+1 &\text{ if } \phi(x,y,t)<0 \\
1&\text{ if } \phi(x,y,t)\geq0 
\end{cases},\quad v=\begin{cases}
-(x-t+0.5) (r^2 -1) +1&\text{ if } \phi(x,y,t)<0 \\
1&\text{ if } \phi(x,y,t)\geq0 
\end{cases},\\
p=\begin{cases}
2-r^2 &\text{ if } \phi(x,y,t)<0 \\
0 &\text{ if } \phi(x,y,t)\geq0 
\end{cases}
\end{gathered}
\end{equation*}
for $r=\sqrt{(x-t+0.5)^2+(y-t+0.5)^2}$. The profile of the solution is shown in Figure \ref{fig:analytic_moving}. The process of checking $\uvec=(1,1)$ on $\Gamma_t$ is easy, and therefore, movement of the interface $\Gamma_t$ is easily found to agree with the velocity $\uvec$. Numerical simulations are conducted from $t=0$ to $t=1$, with material quantities $ \mu^- = 0.01, \mu^+ = 0.1$ and $\rho^- = 0.1, \rho^+ = 1$. The convergence order estimates for the velocity, pressure, and interface location are observed in Table \ref{tab:ns_analytic}, where the error of the interface location is measured based on the error of the level-set function on the grid points of $|\phi_{i,j}|<3\Delta x$. Because of the movement of the interface, the convergences of velocity and pressure are not exactly first-order. Nonetheless, we obtain first-order accuracy for the interface position.

\begin{table}[]
	\caption{Convergence of Example \ref{ex:NS_analytic2}.}
	\label{tab:ns_analytic}
	\begin{center}
		\begin{tabular}{llllllll}
			\hline
			
			resolution &   $L^\infty$ error of $\uvec$  & order     &   $L^\infty$ error of $p$        & order     &  $L^\infty$ error of $\phi$        &  order    \\
			\noalign{\smallskip}\hline\noalign{\smallskip}
			$32\times 32$   & 1.22.E-01 &      & 9.81.E-02 &      & 3.58.E-02 &      \\
			$64\times 64$   & 7.03.E-02 & 0.80 & 5.22.E-02 & 0.91 & 1.66.E-02 & 1.10 \\
			$128\times 128$ & 3.96.E-02 & 0.83 & 3.67.E-02 & 0.51 & 8.32.E-03 & 1.00 \\
			$256\times 256$ & 2.10.E-02 & 0.91 & 1.51.E-02 & 1.28 & 4.20.E-03 & 0.98
		\end{tabular}
	\end{center}
\end{table}
\subsection{Parasitic currents}\label{ex:parasitic}
As a fourth example, we consider a Navier--Stokes equation that accounts for surface tension in the absence of gravity. The initial interface is given as a circle, resulting in an exact solution that has zero velocity, and piece-wise constant pressure, depending on the radius of the circle and surface tension coefficient. This example is also called a parasitic current, which has been tested in \cite{schroeder2014second,theillard2019sharp,sussman2007sharp}. We perform simulations up to $t=0.5$, where the initial interface is given as $\phi= \sqrt{x^2 +y^2}-1$ on the computational domain \(\Omega =[-2,2]^2\). The following values for density, viscosity, and surface tension coefficient, 
\[ \rho^+ =0.1,\quad  \rho^- =1,\quad  \mu^+ =0.01,\quad  \mu^- =0.1,\quad \beta=50,\] are chosen. $L^\infty$ errors of velocity, pressure, and interface location are presented in Table \ref{tab:ns_parasitic}, showing first-order convergence.
\begin{table}[]
	\caption{Convergence of Example \ref{ex:parasitic}.}
	\label{tab:ns_parasitic}
	\begin{center}
		\begin{tabular}{llllllll}
			\hline
			
			resolution &   $L^\infty$ error of $\uvec$  & order     &   $L^\infty$ error of $p$        & order     &  $L^\infty$ error of $\phi$        &  order    \\
			\noalign{\smallskip}\hline\noalign{\smallskip}
			$32\times 32$   & 8.77.E-02 &      & 4.72.E-01 &      & 3.67.E-03 &      \\
			$64\times 64$   & 4.00.E-02 & 1.13 & 2.54.E-01 & 0.89 & 2.10.E-03 & 0.81 \\
			$128\times 128$ & 1.81.E-02 & 1.14 & 1.53.E-01 & 0.73 & 6.62.E-04 & 1.66 \\
			$256\times 256$ & 1.06.E-02 & 0.78 & 7.87.E-02 & 0.96 & 2.54.E-04 & 1.38
		\end{tabular}
	\end{center}
\end{table}

\subsection{Rising bubble}\label{ex:bubble}
Lastly, we simulate rising-bubbles problems involving strong surface tension. For each numerical simulation, we measure the rising velocity, circularity, and relative area loss of the bubble. Each of these quantities are calculated as follows:
\begin{equation*}
\begin{aligned}
\text{rising velocity}&=\frac{\int _\Omega(1- H(\phi))\uvec dx}{\pi r_0^2 }, \quad \text{circularity}&=\frac{2\pi r_0}{ \int_\Omega \delta(\phi) dx}, \quad
\text{relative area loss} & = \frac{\pi r_0^2-\int _\Omega 1- H(\phi) dx }{\pi r_0^2}
\end{aligned}
\end{equation*}
$r_0$ is the initial radius of the bubble, and $H,\delta$ are the numerical Heaviside and delta functions, which are defined as
\begin{equation*}
H (\phi) = \begin{cases}
0  &\text{     }\phi<-\epsilon\\
\frac{1}{2}+\frac{\phi}{2\epsilon}+\frac{1}{2\pi}\sin (\frac{\pi \phi}{\epsilon})  &-\epsilon \leq\phi\leq\epsilon\\
1   &\epsilon<\phi\\
\end{cases}, \quad
\delta (\phi) = \begin{cases}
0  &\text{     }\phi<-\epsilon\\
\frac{1}{2\epsilon}+\frac{1}{2\epsilon}\cos (\frac{\pi \phi}{\epsilon})  &-\epsilon \leq\phi\leq\epsilon\\
0   &\epsilon<\phi\\
\end{cases}.
\end{equation*}
In our numerical experiments, we set $\epsilon=1.5 \Delta x$.
\subsubsection{Small-air-bubble rising} \label{ex:small_air}
\begin{figure}
	\centering
	\subfloat[a][$t=0.02$]{	\includegraphics[width=0.3\textwidth]{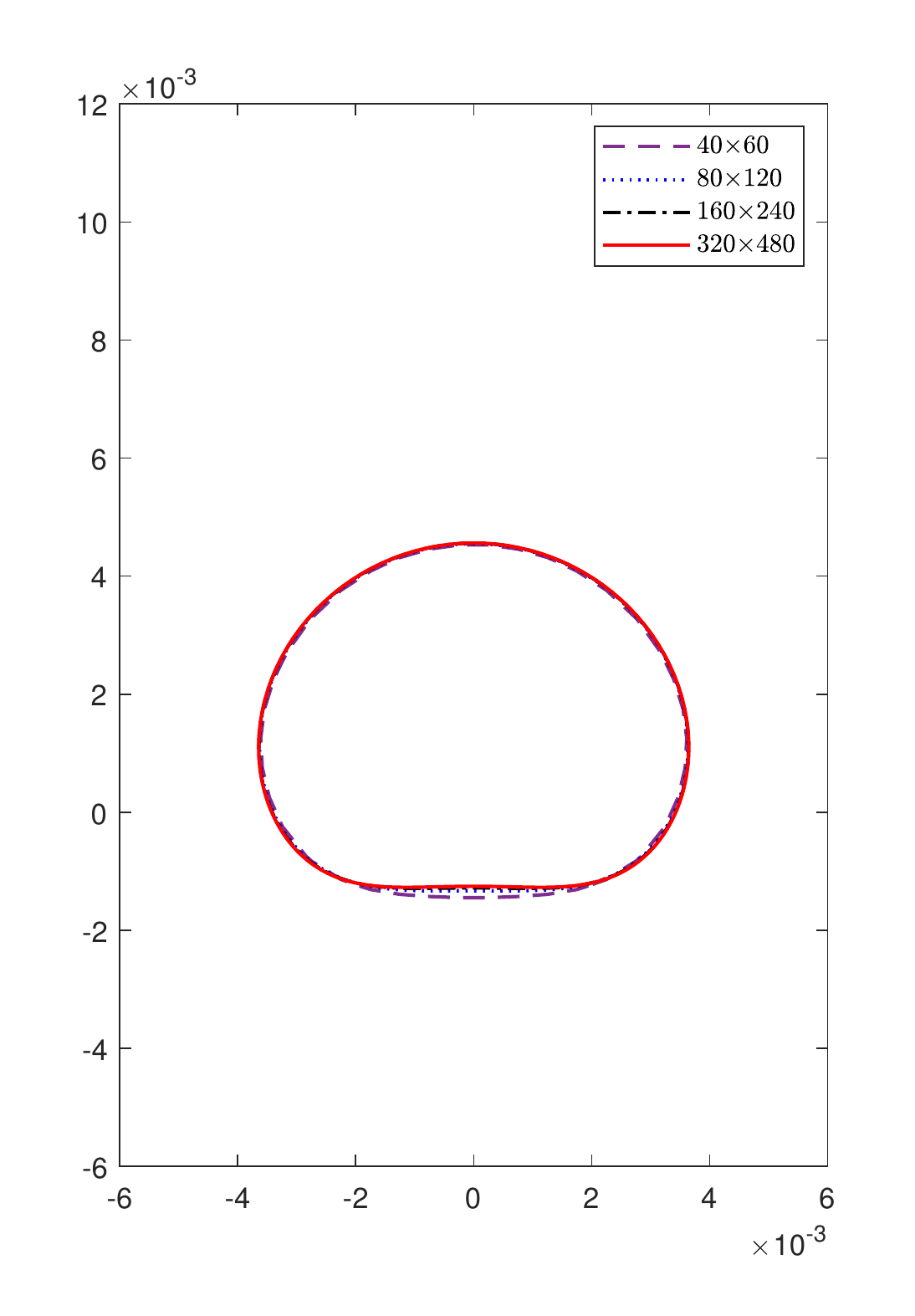}}
	\subfloat[b][$t=0.035$]{\includegraphics[width=0.3\textwidth]{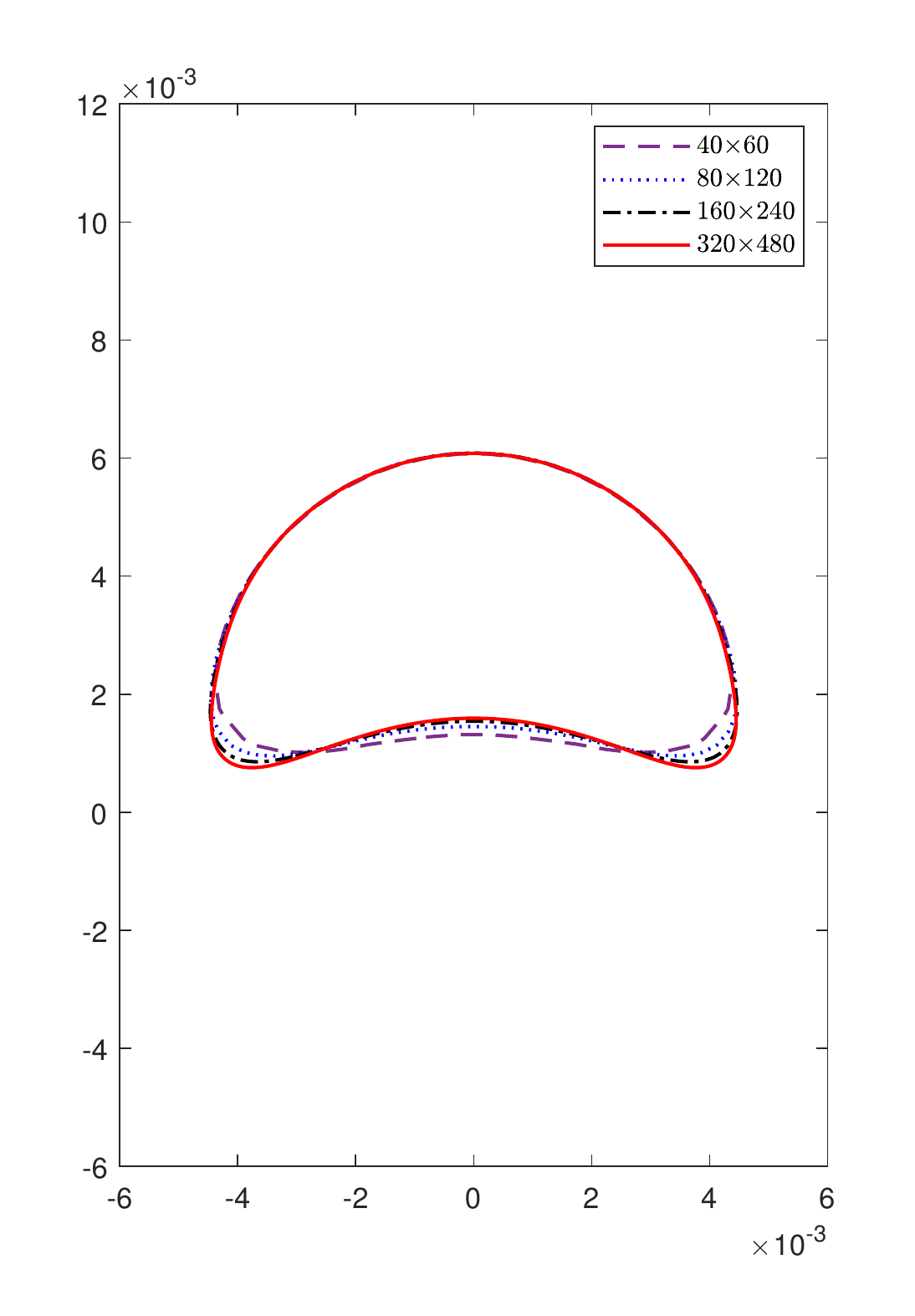}}
	\subfloat[c][$t=0.05$]{\includegraphics[width=0.3\textwidth]{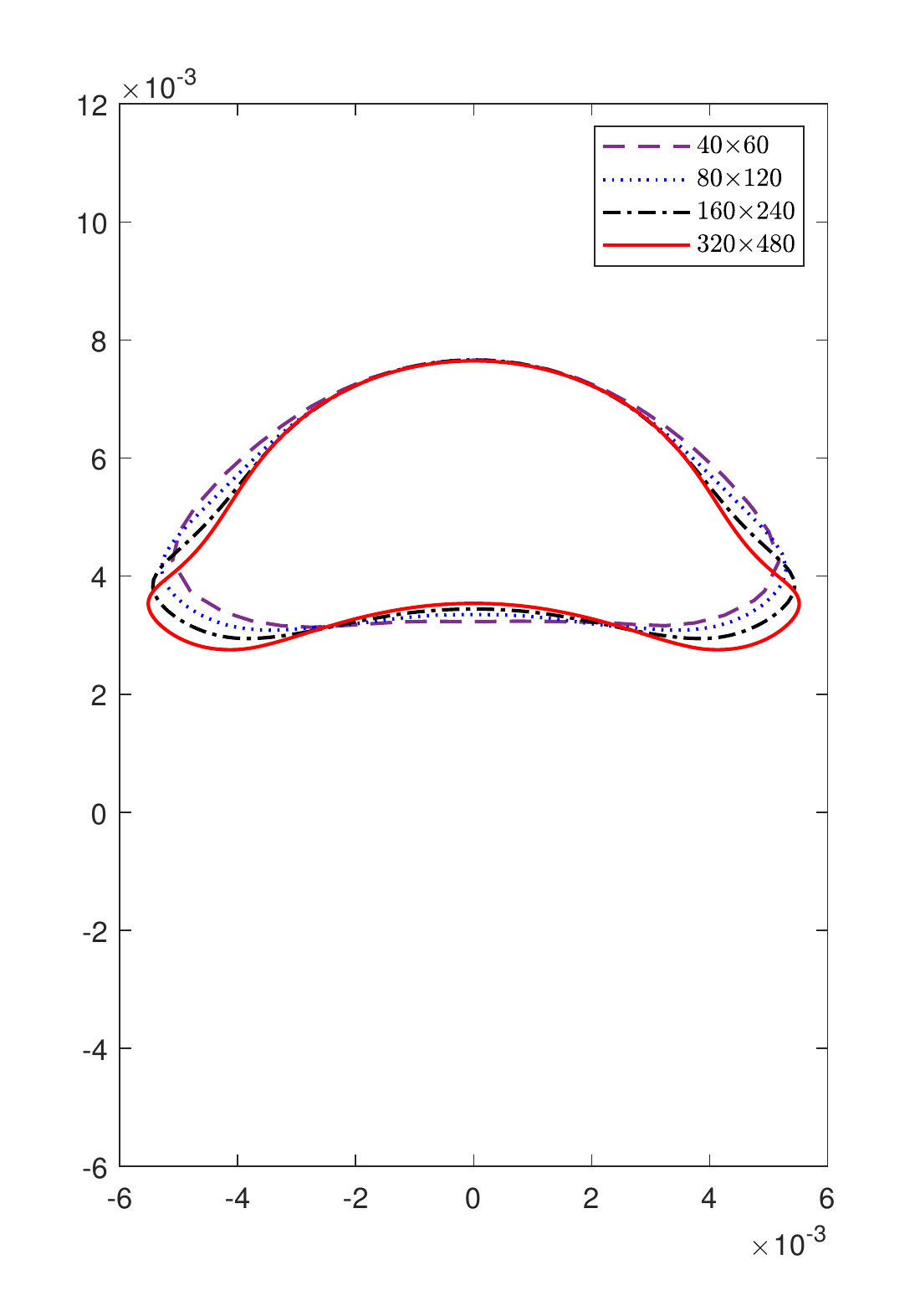}}
	\caption{Visualization of solution for example \ref{ex:small_air}.}
	\label{fig:bubblesol}
\end{figure}

\begin{figure}
	\centering
	\subfloat[a][Rising velocity]{	\includegraphics[width=0.3\textwidth]{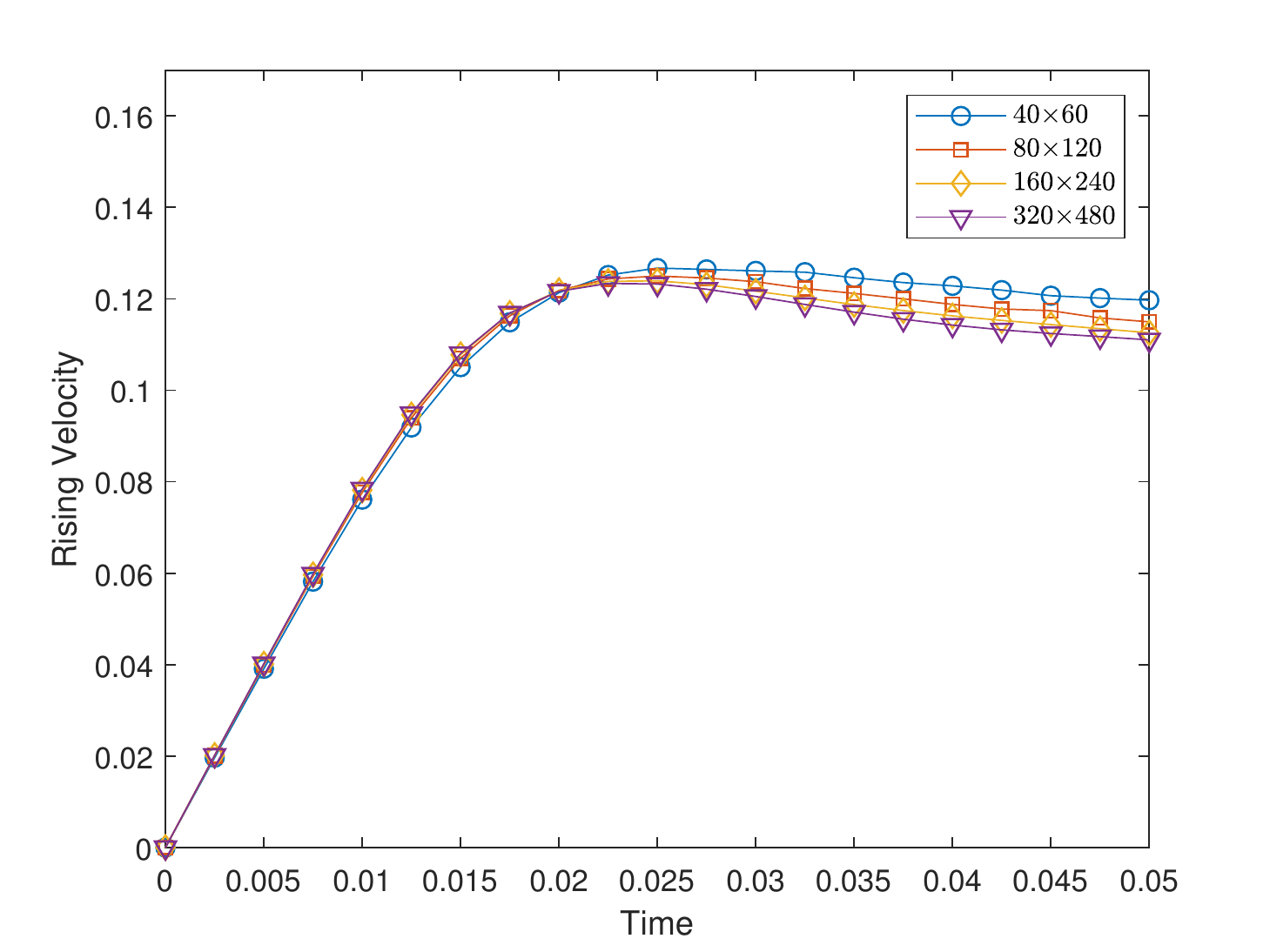}}
	\subfloat[b][Circularity] {\includegraphics[width=0.3\textwidth]{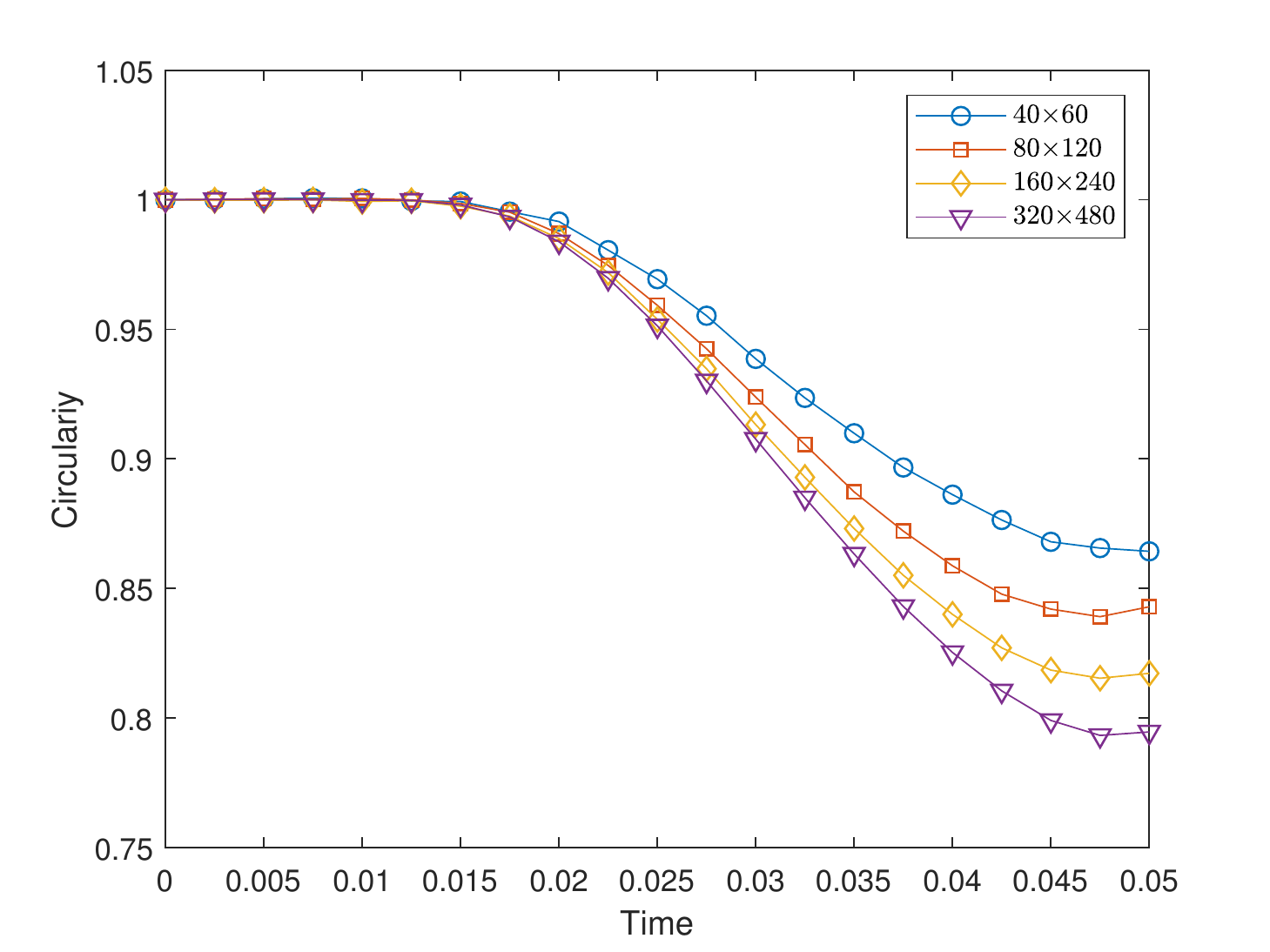}}
	\subfloat[c][Relative area loss]{\includegraphics[width=0.3\textwidth]{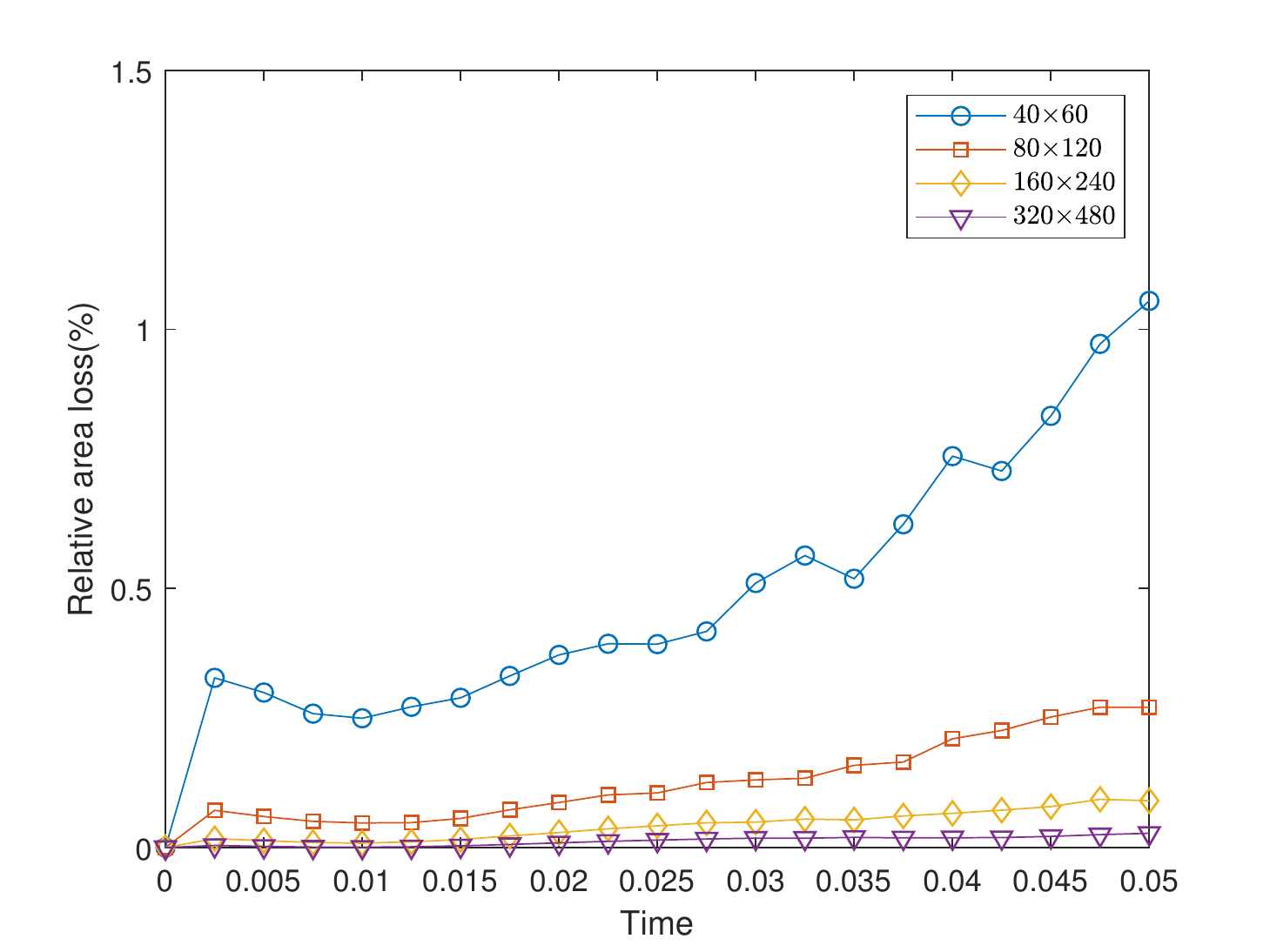}}	\caption{Rising velocity, circularity, and relative area loss for small air bubble in \ref{ex:small_air}.}
	\label{fig:bubble1_quantities}
\end{figure}

First, we simulate an air bubble rising in a tank filled with water, which has been tested in \cite{kang2000boundary}. Density, viscosity, surface tension coefficient, and gravity are set realistically, with the values 
\[\rho^+ =1000,\quad \rho^- =1.226, \quad 	\mu^+ = 1.137 \times 10^{-3}, \quad \mu^- =  1.78 \times 10^{-5},\quad \beta=0.0728, \quad \mathbf{g}= \left(0,-9.8\right).\]
The interface is initialized with level-set function $\phi(x,y,0)=\sqrt{x^2+y^2}-\frac{1}{300}$ on the computational domain $\Omega= [-0.01,0.01]\times [-0.01,0.02]$ with a no-slip boundary condition for the velocities. Calculations are performed in Cartesian grids of sizes $40\times 60,80\times 120,160\times 240$, and $320\times 480$. Visualizations of the air bubble at $t=0.02, 0.035,$ and $0.05$ are shown in Figure \ref{fig:bubblesol}, verifying the convergence of the interface position. Rising velocity, circularity, and relative area loss are presented in Figure \ref{fig:bubble1_quantities}. Relative area losses on the four different grids at $t=0.05$ are $1.36, 0.34, 0.11$, and $0.03\%$, respectively. Compared to the results for GFM\cite{kang2000boundary}, which were $17.23, 5.76, 1.54,$ and $0.0036\%$, the results of our air-bubble-rising experiment demonstrate a significant improvement in terms of area preservation at a low resolution. 

\subsubsection{Two-dimensional bubble-rising benchmark test by Hysing et al. \cite{hysing2009quantitative} } \label{ex:hysing}
\begin{figure}
	\centering
	\subfloat[a][$t=1.5$]{	\includegraphics[width=0.5\textwidth]{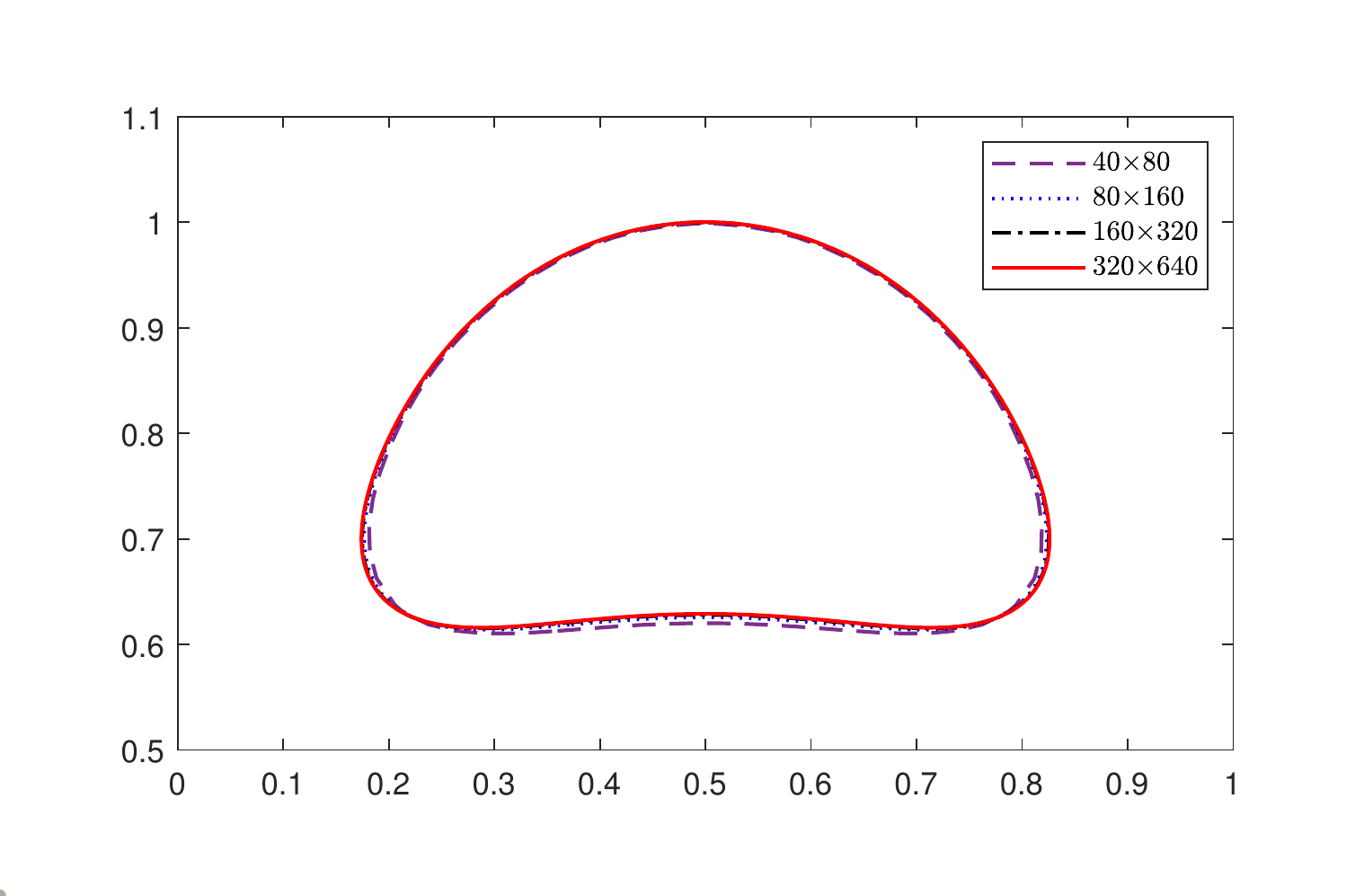}}
	\subfloat[b][$t=1.5$(zoomed)] {\includegraphics[width=0.5\textwidth]{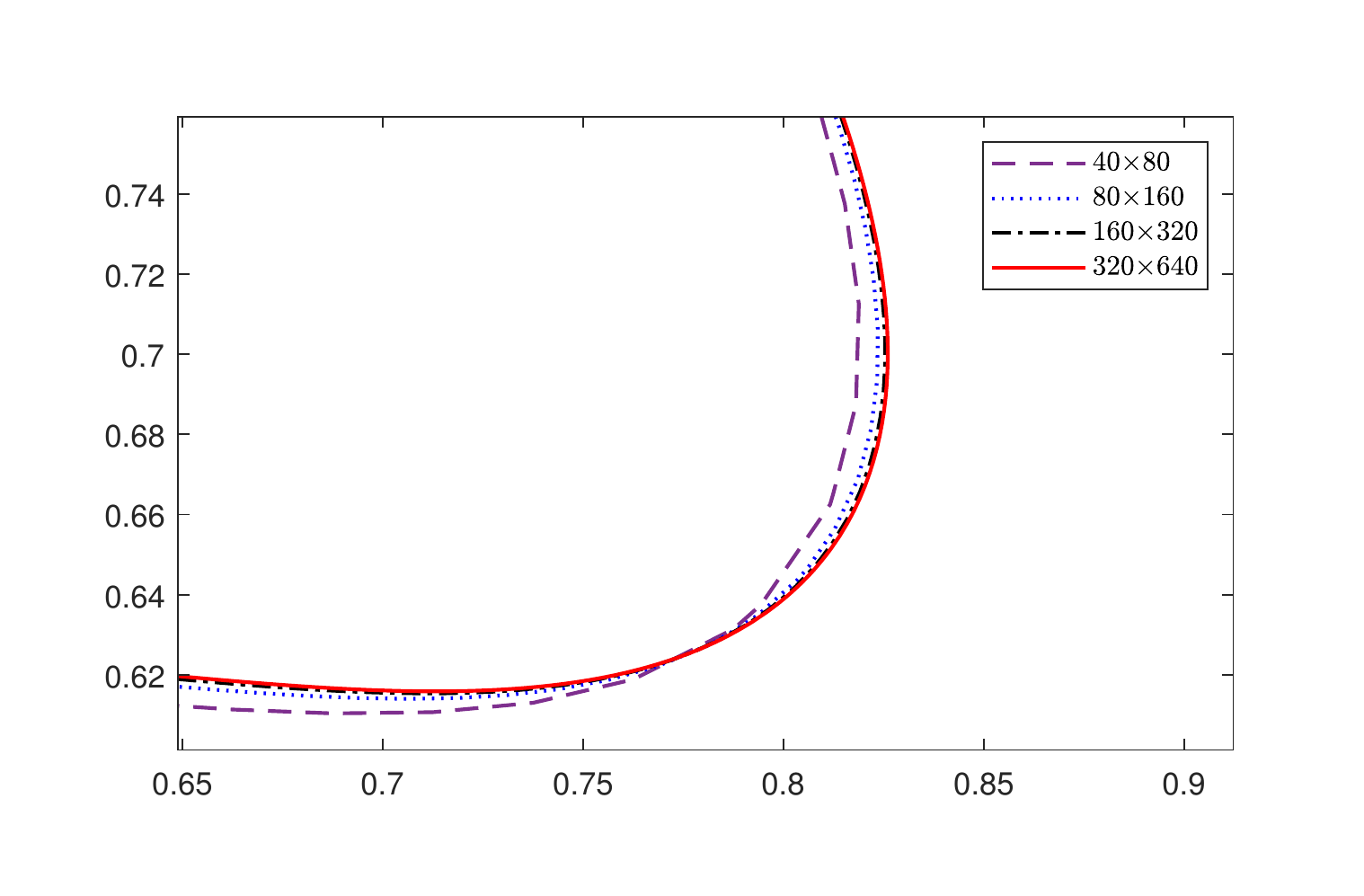}}\\
	\subfloat[c][$t=3$]{\includegraphics[width=0.5\textwidth]{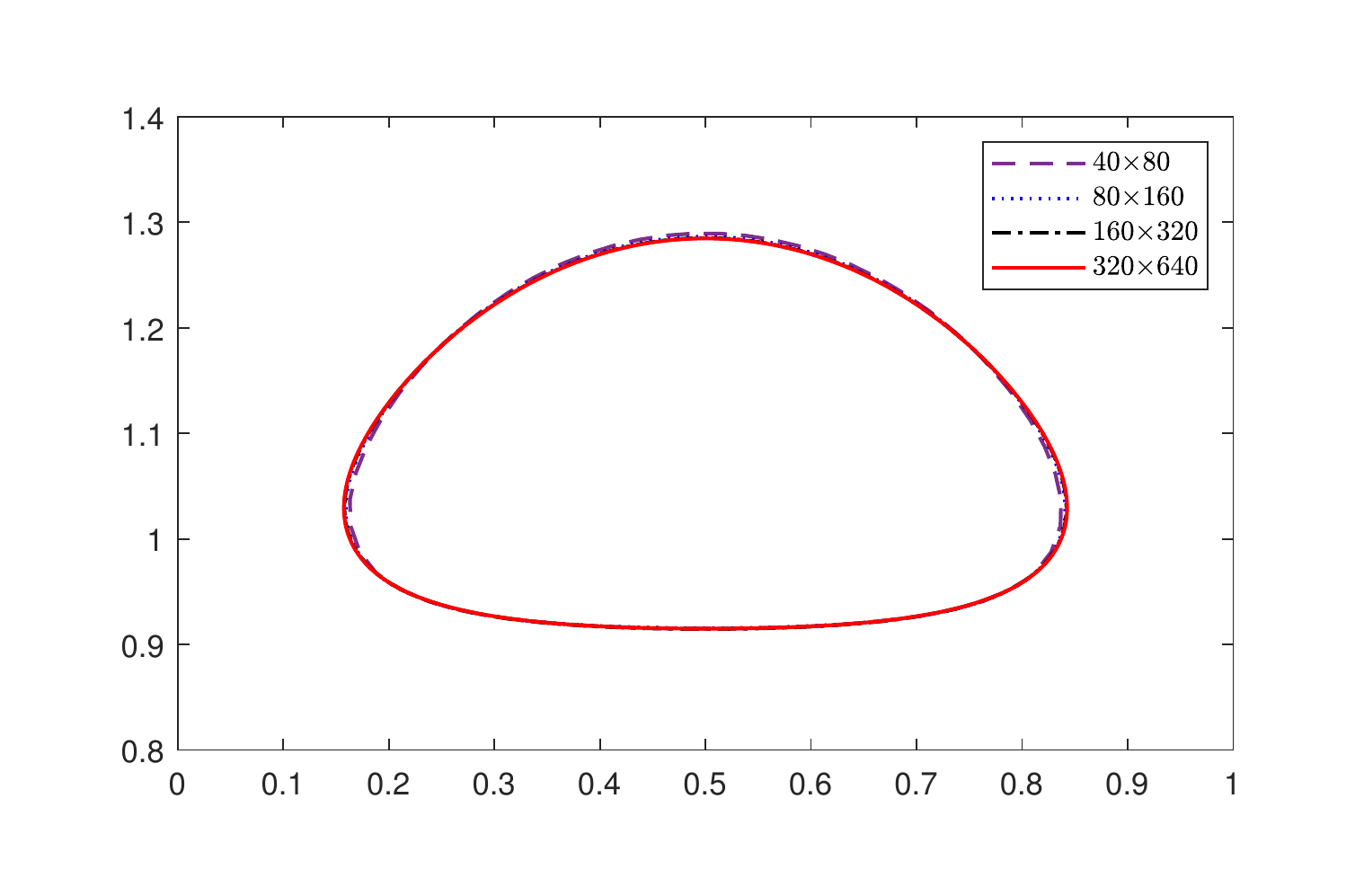}}
	\subfloat[d][$t=3$(zoomed)]{\includegraphics[width=0.5\textwidth]{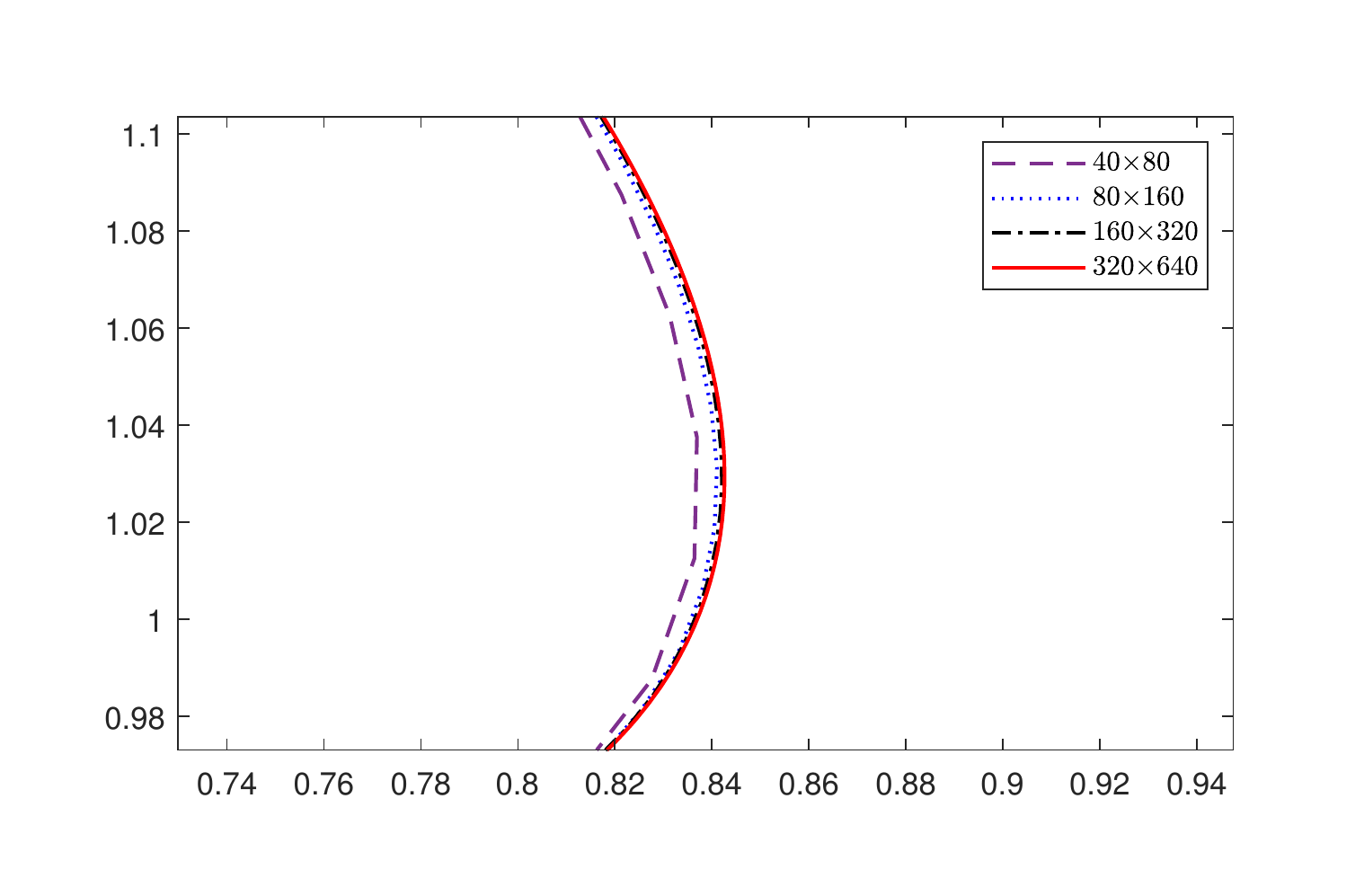}}		\caption{Visualization of solution for example \ref{ex:hysing}.}
	\label{fig:bubble2}
\end{figure}

\begin{figure}
	\centering
	\subfloat[a][Rising velocity]{	\includegraphics[width=0.3\textwidth]{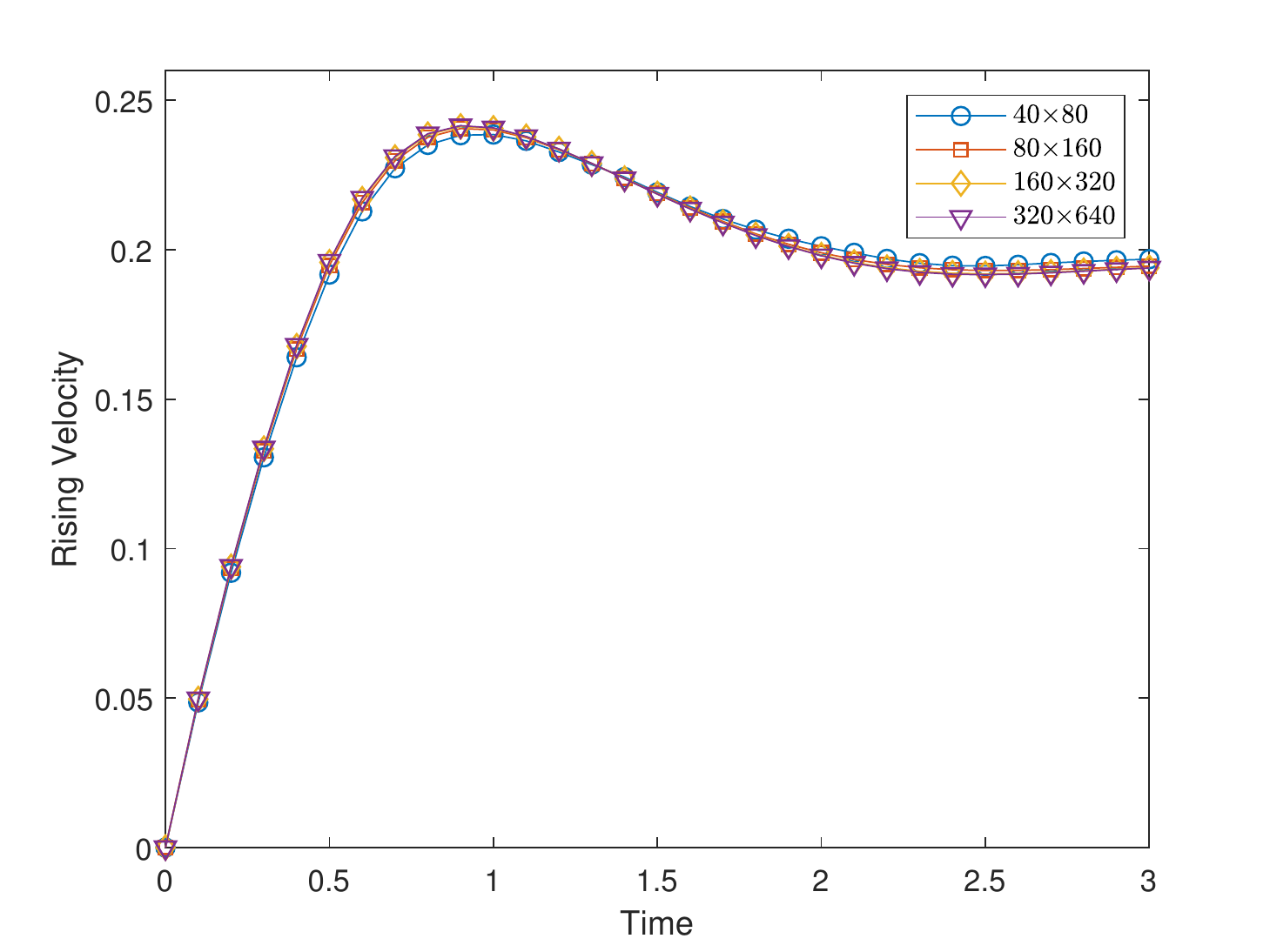}}
	\subfloat[b][Circularity] {\includegraphics[width=0.3\textwidth]{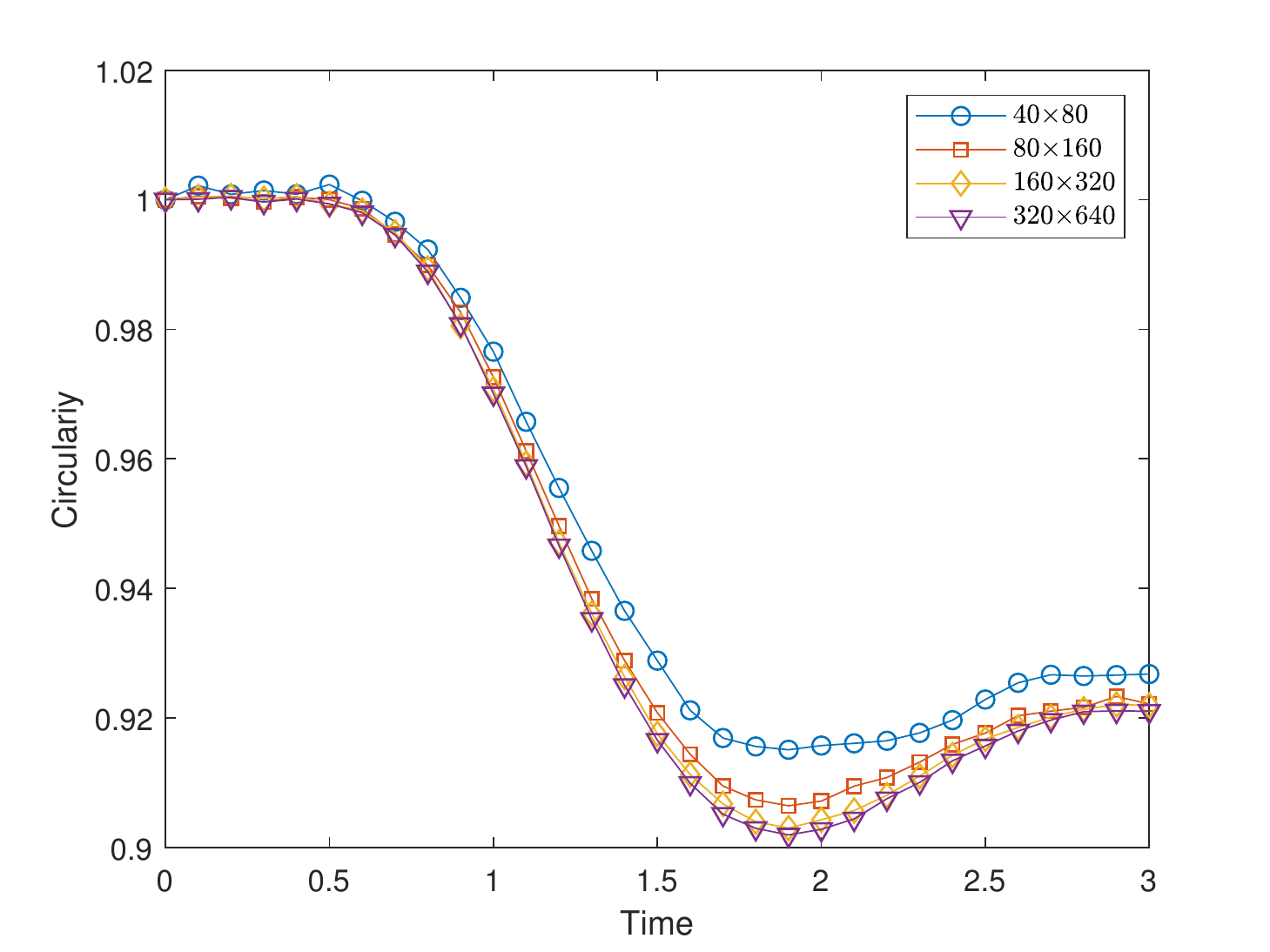}}
	\subfloat[c][Relative area loss]{\includegraphics[width=0.3\textwidth]{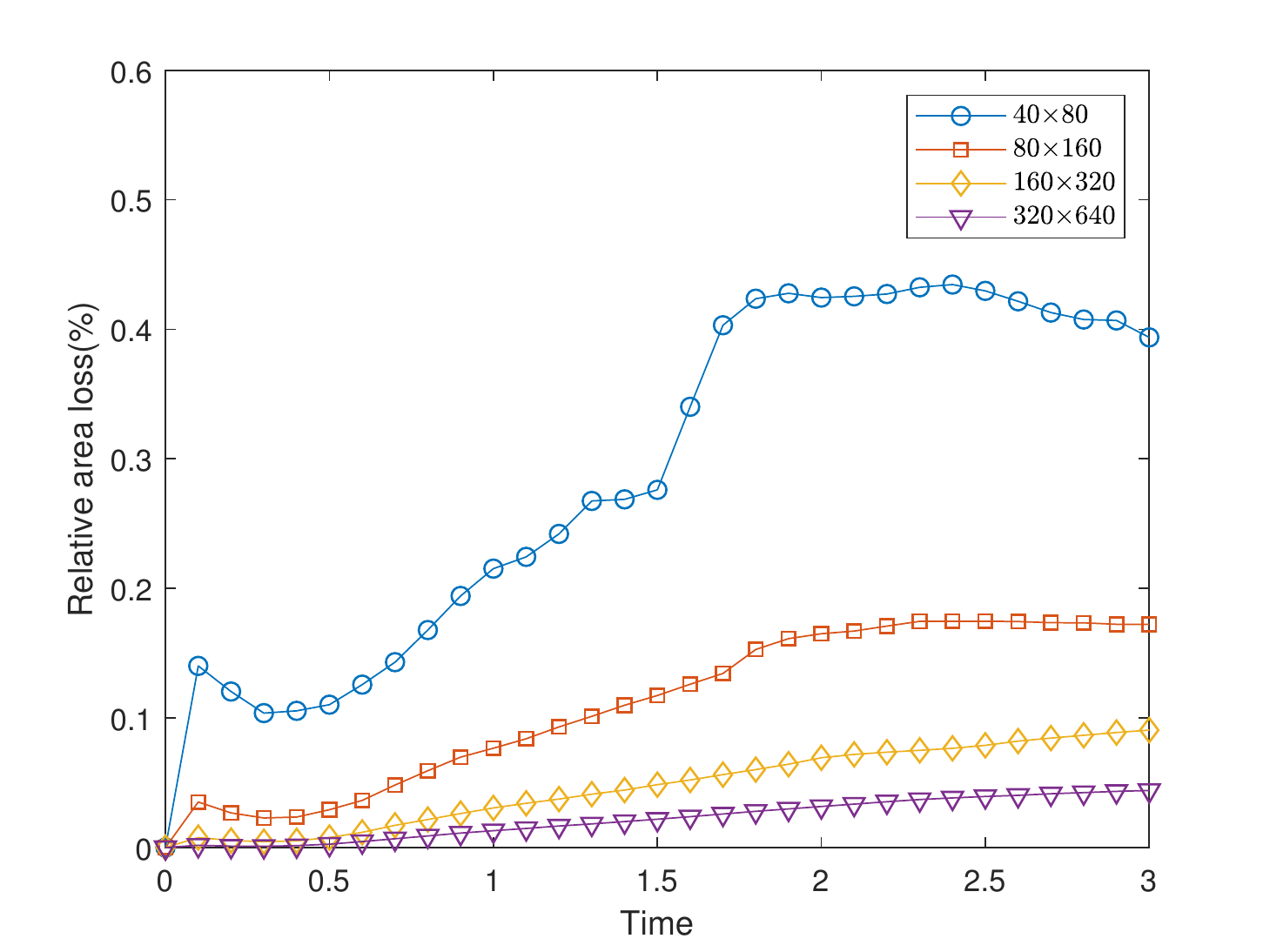}}	\caption{Rising velocity, circularity, and relative area loss for rising bubble in \ref{ex:hysing}.}
	\label{fig:bubble2_quantities}
\end{figure}

We consider case 1 of the benchmark problem proposed in \cite{hysing2009quantitative}. Material quantities are then assigned the values 
\[\rho^+ =1000,\quad \rho^- =100, \quad 	\mu^+ = 10 , \quad \mu^- =  1 ,\quad \beta=24.5, \quad \mathbf{g}= \left(0,-0.98\right).\]
On the computational domain $\Omega= [0,1]\times [0,2]$, the bubble is initialized as a circle centered at \(\left(0.5,1 \right) \) with radius $r_0=0.25$. A no-slip boundary condition is imposed on the horizontal wall, whereas free-slip boundary condition is imposed on the vertical wall. This example is experimented on uniform Cartesian grids with sizes $40\times 80,80\times 160,160\times 320$, and $320\times 640$, up to $t=3$. Visualizations of the bubble at $t=1.5$ and $3$ are shown in Figure \ref{fig:bubble2}. In contrast to \ref{ex:small_air}, little difference exists between the shapes of the bubble as the grid is refined. Nonetheless, one can see the convergence of interface position by zooming in the results. Rising velocity, circularity, and relative area loss are shown in Figure \ref{fig:bubble2_quantities}. The graphs of rising velocity and circularity agree with the results in \cite{hysing2009quantitative}. Convergence of the quantities is also verified.

\section{Conclusion}
In this paper, we presented a sharp capturing method for two-phase incompressible Navier--Stokes equations. We derived jump condition formulas for pressure and velocity gradients to apply with the ghost fluid method. Together with a divergence-free condition, saddle-point systems for velocity and pressure are constructed, allowing the jump condition to be applied on the viscous term and pressure implicitly. The saddle-point system is solved via an iterative method, where each iterative step consists of solving symmetric linear systems and extrapolating interfacial velocities to nearby grid points to correct the jump conditions. The numerical experiment supports the idea that our method is first-order accurate for velocity, pressure, and interface position for analytical solution, and can be applied to practical problems.

Although we proposed a preconditioner for the saddle-point problem, the construction of a more effective preconditioner for the saddle-point problem, and its detailed analysis, remain as prospects for future work. Furthermore, we will consider second-order extension of the proposed method, by applying the jump condition to the convection term and considering jump conditions for gradients of pressure.

	\bibliographystyle{spmpsci} 
	\bibliography{mybib}

\end{document}